\RequirePackage{fix-cm}
\documentclass{svjour3}                     % onecolumn (standard format)
\smartqed  % flush right qed marks, e.g. at end of proof
\usepackage{cite}
\usepackage{amsmath,amssymb,amsfonts}
\usepackage{algorithmic}
\usepackage{graphicx}
\usepackage{textcomp}
\usepackage[table]{xcolor}
\usepackage{longtable}
\usepackage{adjustbox}
\usepackage{float}
\usepackage{booktabs}
\usepackage{makecell}
\usepackage{pdflscape}

\journalname{Empirical Software Engineering}
\newcommand{\RQone}{\emph{What are the eventual connectivity issues observed in Android open-source apps?}}

\usepackage{amssymb}
\usepackage[hidelinks]{hyperref}
\usepackage[plain]{fancyref}
\usepackage{ifdraft}

\usepackage[inline]{enumitem}
\usepackage{xcolor}
\usepackage{xspace}
\usepackage[final]{listings}
\usepackage{acronym}
\usepackage{url}
\usepackage{amsmath}
\usepackage{amssymb}
\usepackage{booktabs} % For formal tables
\usepackage{subfigure}
\usepackage{balance}
\usepackage{dirtree}

\usepackage[ruled]{algorithm2e} % For algorithms

\newcommand{\subparagraph}{}
\usepackage{etoolbox}
\makeatletter
\patchcmd{\@makecaption}
  {\scshape}
  {}
  {}
  {}
\patchcmd{\@makecaption}
  {\\}
  {.\ }
  {}
  {}
\makeatother

\definecolor{OliveGreen}{rgb}{0,0.6,0.3}

\def\fref{\Fref} % treat all \frefs as \Frefs
\renewcommand{\lstlistingname}{Snippet}
\newcommand*{\fancyreflstlabelprefix}{lst}
\newcommand*{\Freflstname}{\lstlistingname}
\newcommand*{\freflstname}{\MakeLowercase{\lstlistingname}}
\Frefformat{vario}{\fancyreflstlabelprefix}%
  {\Freflstname\fancyrefdefaultspacing#1#3}
\frefformat{vario}{\fancyreflstlabelprefix}%
  {\freflstname\fancyrefdefaultspacing#1#3}
\Frefformat{plain}{\fancyreflstlabelprefix}%
  {\Freflstname\fancyrefdefaultspacing#1}
\frefformat{plain}{\fancyreflstlabelprefix}%
  {\freflstname\fancyrefdefaultspacing#1}

\renewcommand{\tablename}{Table}  
  
\newcommand*{\fancyreflnlabelprefix}{ln}
\newcommand*{\Freflnname}{Line}
\newcommand*{\freflnname}{\MakeLowercase{\Freflnname}}
\Frefformat{vario}{\fancyreflnlabelprefix}%
  {\Freflnname\fancyrefdefaultspacing#1#3}
\frefformat{vario}{\fancyreflnlabelprefix}%
  {\freflnname\fancyrefdefaultspacing#1#3}
\Frefformat{plain}{\fancyreflnlabelprefix}%
  {\Freflnname\fancyrefdefaultspacing#1}
\frefformat{plain}{\fancyreflnlabelprefix}%
  {\freflnname\fancyrefdefaultspacing#1}

\lstdefinelanguage{JavaScript}{
keywords={typeof, new, true, false, catch, function, return, null, catch, switch, var, if, in, for, while, do, else, case, break, throw, this, instanceof},
keywordstyle=\color{purple}\bfseries,
ndkeywords={},
ndkeywordstyle=\color{blue}\bfseries,
identifierstyle=\color{black},
sensitive=false,
comment=[l]{//},
morecomment=[s]{/*}{*/},
commentstyle=\color{OliveGreen}\ttfamily,
stringstyle=\color{OliveGreen}\ttfamily,
morestring=[b]',
morestring=[b]"
}
\usepackage{color}
\definecolor{gray97}{gray}{.97}
\definecolor{gray90}{gray}{.90}
\definecolor{gray75}{gray}{.75}
\definecolor{gray45}{gray}{.45}
\definecolor{codegreen}{rgb}{0,0.6,0}
\definecolor{codered}{rgb}{0.6,0,0}
\definecolor{codegray}{rgb}{0.5,0.5,0.5}
\definecolor{codepurple}{rgb}{0.58,0,0.82}
\lstdefinestyle{floating}{%
  frame=none,
  float=htb,
  captionpos=b
}

\lstdefinestyle{ctxtraits}
 {language=JavaScript,
  frame=lines,
  showstringspaces=false,
  keywordstyle=\tt\bf,
  tabsize=3,
  style=floating,
  morekeywords={Trait, cop, Context, activate, deactivate, adapt, addObjectPolicy, manager}
}

\lstnewenvironment{ctxtraits}[1][]
 {\lstset{style=ctxtraits,#1}}{}

\newcommand{\eg}{\emph{e.g.,}\xspace}
\newcommand{\ie}{\emph{i.e.,}\xspace}
\newcommand{\etal}{\emph{et al.}\xspace}

\newcommand{\secref}[1]{Section~\ref{#1}\xspace}
\newcommand{\figref}[1]{Fig.~\ref{#1}\xspace}
\newcommand{\listref}[1]{Snippet~\ref{#1}\xspace}

\newcommand{\etc}{\textit{etc.}\xspace}

\newcommand{\sfigref}[1]{Subfig.~\ref{#1}\xspace}

\definecolor{author}{rgb}{.5, .5, .5}
\definecolor{comment}{rgb}{.1, .0, .9}
\definecolor{note}{rgb}{.9, .4, .0}
\definecolor{idea}{rgb}{.1, .7, .0}
\definecolor{missing}{rgb}{.9, .1, .0}
\definecolor{deleteme}{rgb}{.9, .1, .0}

\acrodef{APK}{Android Application Package}
\acrodef{ECn}{Eventual Connectivity}

\begin{document}

\title{Studying Eventual Connectivity Issues in Android Apps}

\author{Camilo Escobar-Vel\'asquez \and Alejandro Mazuera-Rozo \and Claudia Bedoya \and Michael Osorio-Ria\~no\and Mario Linares-V\'asquez \and Gabriele Bavota
}

\authorrunning{Escobar-Vel\'asquez \etal} % if too long for running head

\institute{Camilo Escobar-Vel\'asquez \at
              Universidad de los Andes, Bogot\'a, Colombia \\
              \email{ca.escobar2434@uniandes.edu.co}
           \and
           Alejandro Mazuera-Rozo \at
              Universit\`a della Svizzera italiana, Lugano, Switzerland\\
              Universidad de los Andes, Bogot\'a, Colombia\\
              \email{alejandro.mazuera.rozo@usi.ch}
           \and
            Claudia Bedoya \at
           Universidad de los Andes, Bogot\'a, Colombia\\
           \email{cd.bedoya212@uniandes.edu.co}
           \and
           Michael Osorio-Ria\~no \at
           Universidad de los Andes, Bogot\'a, Colombia\\
           \email{ms.osorio@uniandes.edu.co}
           \and
           Mario Linares-V\'asquez \at
           Universidad de los Andes, Bogot\'a, Colombia\\
           \email{m.linaresv@uniandes.edu.co}
           \and
           Gabriele Bavota \at
           Universit\`a della Svizzera italiana, Lugano, Switzerland\\
           \email{gabriele.bavota@usi.ch}
}

\date{Received: \today \ / Accepted: date}

\maketitle

\begin{abstract}
Mobile apps have become indispensable for daily life, not only for individuals but also for companies/organizations that offer their services digitally. Inherited by the mobility of devices, there are no limitations regarding the locations or conditions in which apps are being used. For example, apps can be used where no internet connection is available. Therefore, \textit{offline-first} is a highly desired quality of mobile apps. Accordingly, inappropriate handling of connectivity issues and miss-implementation of good practices lead to bugs and crashes occurrences that reduce the confidence of users on the apps' quality.  In this paper, we present the first study on \textit{Eventual Connectivity} (ECn) issues exhibited by Android apps, by manually inspecting 971 scenarios related to 50 open-source apps. We found 304 instances of ECn issues (6 issues per app, on average) that we organized in a taxonomy of 10 categories. We found that the majority of ECn issues are related to the use of messages not providing correct information to the user about the connectivity status and to the improper use of external libraries/apps to which the check of the connectivity status is delegated. Based on our findings, we distill a list of lessons learned for both practitioners and researchers, indicating directions for future work.
\end{abstract}

\section{Introduction}

Mobile apps have become an indispensable tool for daily activities. There are no limitations regarding the locations or conditions in which mobile devices are used, and  \textit{offline-first} practice is a highly desired quality of mobile apps. However, despite high-speed access to the internet is more and more common worldwide and ``data plans'' are  more accessible/cheap in terms of costs, there are still complex connectivity scenarios, such as locations with zero/unreliable connection. Therefore, inappropriate handling of connectivity issues may lead to bugs and crashes that negatively affect the user experience when an app is used under \textit{\acf{ECn}} scenarios.

\textit{Offline first} practices -- i.e., programming practices that allow an app to provide network-enabled features without internet  access -- have been initially promoted for web applications (in particular progressive web apps), and are motivated by the need for providing a better user experience even when there is no connectivity \cite{Google:OfflineFirst, Archibald:2018}. Several specific implementation practices are well known in the web development community, such as using app shells, local caching in the browser, bundle network requests when users are offline, and connectivity awareness. These practices have also been transferred to the mobile app domain and complemented with app-specific ``guidelines/practices" proposed by practitioners \cite{Machado:2015} and Google developers \cite{Google:2017}. An example of those guidelines is offline data-synchronization with backend services via APIs (\eg local caching enabled by Firebase).

As of today, there is no empirical study analyzing bugs/crashes in Android apps that are caused by connectivity issues. As a consequence, it is unknown the prevalence of those issues and their impact on the quality as perceived by users. Indeed, state-of-the-art studies are mostly devoted to (i) network traffic characterization \cite{Falaki2010, rao2011}, (ii) its security and privacy implications \cite{Kuzuno2013, Song2015}, and (iii) possible exploitable scenarios addressed by malicious agents \cite{Shabtai2011}. Previous studies have indeed focused on cataloging bugs/crashes for mobile apps in general \cite{Khalid:IEEE15},  and bugs/crashes specific to quality attributes such as performance \cite{Guo:2013:CDR:3107656.3107706,Liu:2014:CDP:2568225.2568229,7332486,Lin:2014:RCA:2635868.2635903,Mazuera-Rozo:ESEM19}, security \cite{Agrawal:2013:MMI:2465478.2465497,Allix:MSR16,Zhou:SSP12},  behavioral/GUI inconsistency \cite{7381838,Ali:2017:SAD:3104086.3104099,8115644,Mercado:2016:ICD:2993259.2993268}, and energy consumption \cite{Pathak:HotNets11,Pathak:MobiSys12,Zhang:MOBS13,Liu:PerCom13,Linares-Vasquez:MSR14}.

Moreover, there is no tool available for detecting this type of bugs statically or dynamically. The closest available approaches/tools for automated detection of crashes related to the lack of connectivity are CrashScope \cite{Moran:ICSE17,Moran:ICST16},  Thor \cite{Adamsen:ISSTA15}, and Caiipa \cite{caiipa2014}. Those approaches systematically explore an app, generate events like turning-off WiFi, and look for crashes (\ie the application stops). However, as we will show in this work, crashes caused by a lack of connectivity  only represent a subset of the connectivity issues that affect mobile apps.

Despite the lack of literature strictly related to connectivity issues, the latter are prevalent in Android apps. Indeed, as a preliminary step towards the study we present in this work, we checked whether connectivity-related issues were discussed in the issue trackers of open source Android apps hosted on GitHub. By mining issues from 3,256 apps used in previous work \cite{Geiger2018, Coppola2019} we collected $\sim$219k issues of which 11,350 --- belonging to 943 apps --- matched in their title or description with the keywords \textit{offline}, or \emph{connectivity}, that we used as a mechanism to identify the presence of connectivity-related issues. Knowing that false positives are likely to be retrieved in this way, the first two authors manually inspected 400 instances each to verify whether they were true positives (i.e., reports actually related to connectivity-issues) or false positives; 400 instances ensure a significance interval of $\pm5\%$ with a confidence level of 95\%. After solving 81 conflicts through an open discussion, they agreed on 213 true positives (53.2\%). Being conservative, and assuming a level of precision of 50\% for the employed keyword-based heuristic, this suggests that $\sim$5.6k connectivity-related issues could be present in the issue trackers of the mined apps. More detailed information regarding this analysis can be found on our online appendix \cite{onlineAppendix}.
 
In this work we present the very first study on \textit{Eventual Connectivity (ECn)} issues exhibited by Android apps \textit{in-the-wild}. Our study aims at building the empirical foundations needed to (i) increase practitioners' awareness about the prevalence of these issues and how they manifest in Android apps; and (ii) build techniques and tools aimed at automatically detecting the \textit{ECn} issues we document. Compared to other studies (focused on other type of bugs) that followed a mining-based strategy over code repositories and online markets, we preferred an inspection-based approach to (i)  avoid any of the imprecisions and limitations of  analyzing user reviews \cite{Chen:ICSE14,Villarroel:ICSE16,Palomba:ICSME15,Palomba:JSS18}, and (ii) have detailed information of the conditions that triggered the bugs in the apps. In particular, we manually executed and inspected 50 open source Android apps and we build a catalog of bad practices/issues that are exhibited by those apps in ECn issues. The execution is based on a total of 971 scenarios we designed for the 50 apps; on average, each scenario has 3.5 steps (i.e., interactions with the app).  We found 320 instances of \textit{\ac{ECn}} issues that we grouped into a taxonomy of 10 categories. In addition, we contribute an online appendix~\cite{onlineAppendix} that describes the cause of the identified issues with videos, execution steps and code snippets. Our results and online resources can be used by researchers and practitioners to (i) create approaches for the automated detection of these issues, (ii) being aware of testing cases that should be included into the quality assurance processes of mobile apps, and, in general, (iii) avoid the issues reported in our taxonomy.

\textit{Paper organization.} \secref{sec:std} presents the design of our study, detailing the selection of the context and the process used to test the apps. \secref{sec:results} discusses the achieved results and presents the taxonomy of eventual connectivity issues in Android apps that we built. \secref{section:relatedwork} presents the literature related to different quality aspects of Android apps (e.g., security, performance), while \secref{sec:threats} reports the threats to the validity of our findings. Finally, \secref{sec:conclusions} discusses the learned lessons and our future work.

\section{Study design}
\label{sec:std}

Despite mobile apps widely rely on connections to back-end services/resources via WiFi or cell network, there is no previous work that analyzes the practices followed by Android developers. Therefore, we carried out an empirical study in which we manually analyzed 50 open source Android apps to build a taxonomy and an online catalogue of the most common issues/bad practices in Android apps that are related to eventual connectivity. We relied on open source apps for the analysis because we were interested in the coding practices followed by developers, thus requiring access to the code. The analysis is based on the execution of the apps and the manual localization of the issues in the code. For the 50 apps, we analyzed 971 scenarios, 320 of which allowed to spot issues related to eventual connectivity. The tagging process for  building  the taxonomy was performed by three authors, who categorized each spotted issue and organized them into a taxonomy. The generated taxonomy was revised by the remaining authors. That taxonomy was the foundation for answering the following  question:

\begin{center}
	{\bf RQ$_{1}$:} \RQone
	\vspace{-2px}
\end{center}

With this research question, our goal is to find and classify eventual connectivity issues in a group of open-source Android apps. Also, we intend to understand how these issues affect user-experience and what are the issues in the code causing them. To this aim, we (i) test the apps directly on Android devices, and (ii) check the source code repositories. Additionally, to ease the discussion of the ECn bugs, we also link them to criteria described in the ISO/IEC 25010 quality model, indicating the quality attributes on which each type of bug has an impact. The answer to this question is expected to help developers and researchers to prevent eventual connectivity issues in apps and to develop new tools for automatic identification of \acs{ECn} issues.

In the rest of this section, we describe the open-source mobile apps used for this study and how we selected them. Then, we present the manual analysis we followed to identify/classify the eventual connectivity issues.

\subsection{Context Selection}
\label{sec:study-context}
\begin{figure}[t]
	\begin{center}
		\includegraphics[scale=0.6]{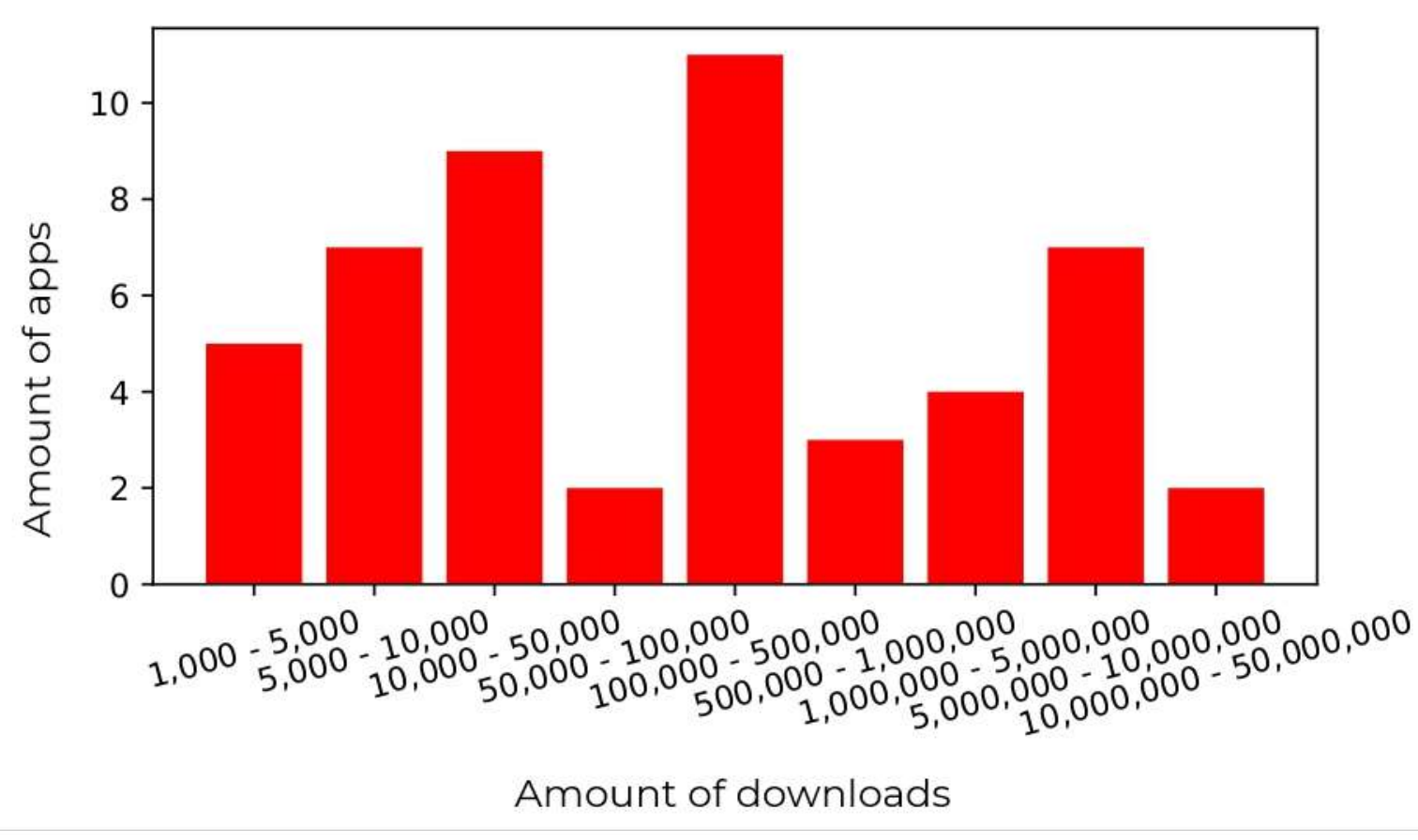}
		\vspace{-0.1cm}
		\caption{Frequency of apps (histogram) per downloads range.}
		\label{fig:apps-downloads}
	\end{center}
\end{figure}

\begin{figure}[t]
	\begin{center}
		\includegraphics[scale=0.5]{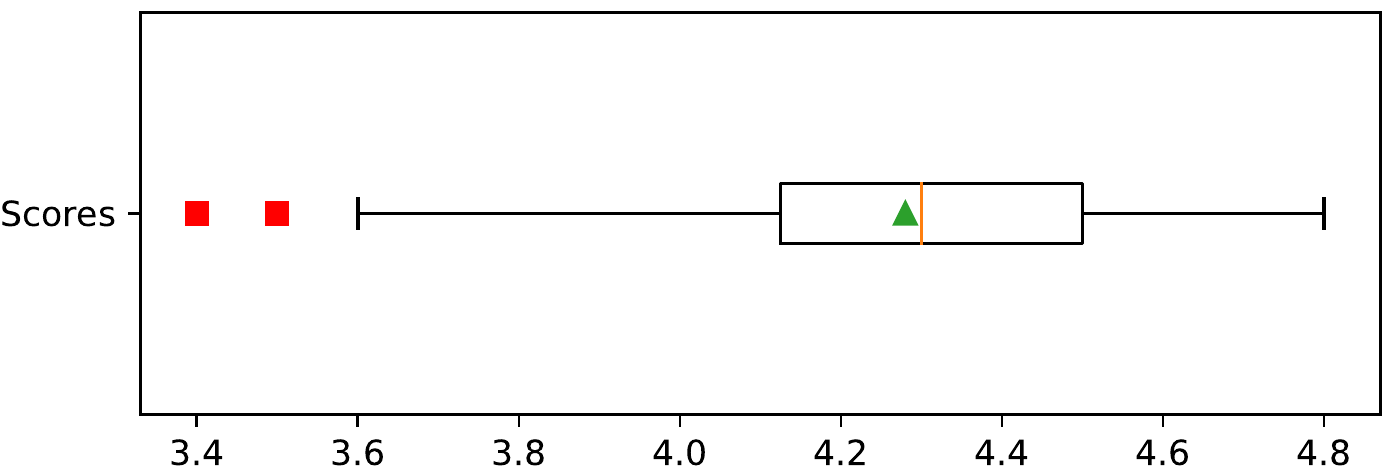}
			\vspace{-0.1cm}
		\caption{Average rating distribution (box plot) of the 50 analyzed Android open source apps. The mean is reported as a triangle and the outliers as squares.}
		\label{fig:apps-rating}
	\end{center}
\end{figure}

\begin{figure}[t]
	\begin{center}
		\includegraphics[scale=0.41]{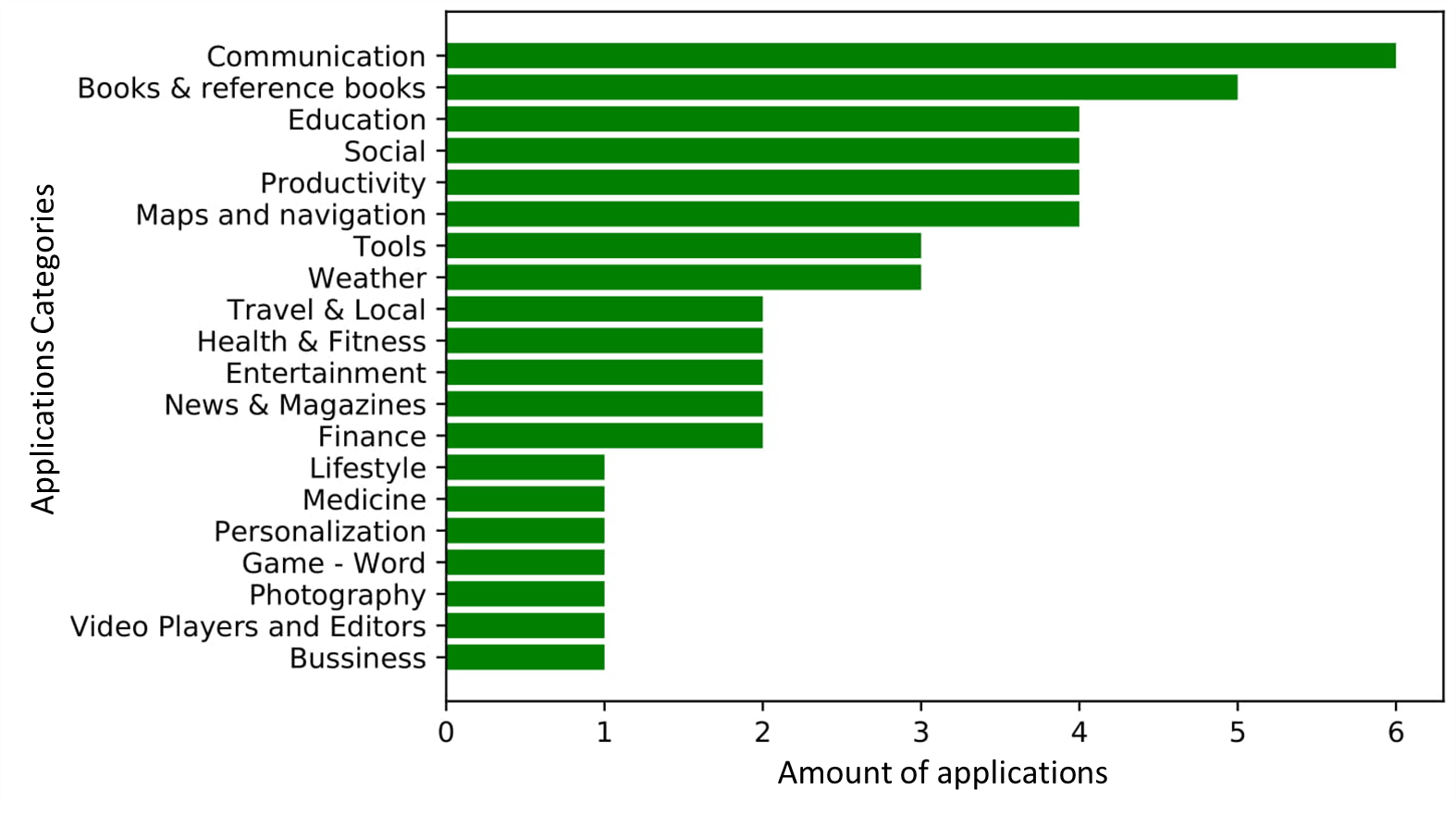}
		\caption{Distribution of number of apps per category.}
		\label{fig:apps-categories}
	\end{center}
\end{figure}

To answer \textbf{RQ$_1$}, we targeted the selection of 50 popular open-source Android apps with features relying on Internet connection. The limit to a maximum of 50 apps was defined due to the expensive manual process adopted in our study to test the apps (details follow). The apps were manually selected from publicly available lists of open source Android apps: (i) the Wikipedia list, featuring 90 free/open source Android apps \cite{wikipedia01}, and (ii) a GitHub repository listing over 300 of open source Android apps \cite{github01}. We looked for apps meeting the following criteria:

\begin{enumerate}
	\item \textit{To be a native Android application (built in Java or Kotlin):} we focused our study on native Android apps excluding hybrid apps, since they may be subject to different types of connectivity issues and we wanted to keep our analysis cohesive. Three authors manually inspected the code repositories of the apps by checking their manifest, the programming language and the code in the project to ensure that the selected apps were not hybrid.
	\item \textit{To be an open-source app available on Google Play:} for this study, we needed access to the source code of the evaluated applications, since we were interested in analyzing code snippets exhibiting issues. In addition, the app should be publicly available on Google Play to ensure that the analyzed apps are neither a library nor a class/toy project.
	\item \textit{To be a popular app:} to guarantee we are studying eventual connectivity behaviors on real apps used by a large amount of people, we focused on popular apps in the market. We consider an application to be popular if (i) it has at least 1,000  downloads and (ii) its average rating is above 3.0. \figref{fig:apps-downloads} depicts the number of apps per downloads range. The download ranges are defined by the Google Play website. The number of downloads is reported as a single number (\eg 500) when it is lower than 1,000, otherwise, the number is reported as a range \eg 10,000 - 50,000. \figref{fig:apps-rating} depicts the distribution of average rating for the 50 apps.
	\item \textit{Covering a diverse set of categories on the Google Play store:} We selected apps covering 20 different categories from the Google Play store, as depicted in \figref{fig:apps-categories}.
	\item \textit{To have functionalities relying on  or network connection:} as we wanted to study issues related to eventual connectivity, we needed our selected apps to have features relying on Internet or network connections. We manually checked this criterion by looking for network-related features in the apps' description (e.g., README file, shared screenshots) and permissions definition.
\end{enumerate}

The list of 50 selected apps is presented in Table ~\ref{tab:one}. Our rationale for limiting the study to 50 applications is based on the amount of effort required to manually test the apps (details about the testing process are reported in the following). The column ``Date'' corresponds to the date in which the information and APKs were downloaded. The ``Downloads interval'', ``Average Rating'',  ``Version'', and the ``Category'' information of each app was collected from the Google Play Store. The values presented in the column ``Version'' refer to the latest version of the artifacts available on the market when we collected the information. As we can see, the set of apps we selected cover a total of 20 different categories, ensuring some diversity of the selected apps. Still, we do not claim that our dataset is representative of all native Android apps, since this would require to run our analyses on a much larger number of apps.

\begin{table}%
	\caption{Analyzed applications.}%To be continued in \tabref{tab:two}}
	\label{tab:one}
	\begin{minipage}{\columnwidth}
		\begin{center}
			\resizebox{0.8\textwidth}{!}{
				\begin{tabular}{m{0.3cm}lllllll}
					\toprule
					
					\thead{ID} & \thead{Application} & \thead{Category} & \thead{Downloads} & \thead{Average  rating} &  \thead{Version} &  \thead{Revision date} \\
					\midrule
					1 & QKSMS & Communication & 100,000 - 500,000 & 4.1 & v2.7.3 & 13/07/2017 \\
					2 & Wire & Communication & 1,000,000 - 5,000,000 & 4.1 & 2.38.352 & 13/08/2017 \\
					3 & Surespot & Social & 100,000 - 500,000 & 4.2 & 70 & 23/08/2017 \\
					4 & Galaxy Zoo & Education & 5,000 - 10,000 & 4.6 & 1.69 & 02/08/2017 \\
					5 & Fanfiction reader & \begin{tabular}[c]{@{}l@{}}Books\\   and reference books\end{tabular} & 100,000 - 500,000 & 4.3 & 1.51 & 20/08/2017 \\
					6 & PressureNet & Weather & 100,000 - 500,000 & 4 & 5.1.10 & 02/08/2017 \\
					7 & AnntenaPod & \begin{tabular}[c]{@{}l@{}}Video\\   Players and Editors\end{tabular} & 100,000 - 500,000 & 4.6 & 1.6.2.3 & 02/08/2017 \\
					8 & iFixit & \begin{tabular}[c]{@{}l@{}}Books\\   and reference books\end{tabular} & 1,000,000 - 5,000,000 & 4.3 & 2.9.2 & 03/08/2017 \\
					9 & DuckDuckGo & \begin{tabular}[c]{@{}l@{}}Books\\   and reference books\end{tabular} & 1,000,000 - 5,000,000 & 4.4 & 3.1.1(101) & 20/08/2017 \\
					10 & OsmAnd & \begin{tabular}[c]{@{}l@{}}Maps\\   and navigation\end{tabular} & \begin{tabular}[c]{@{}l@{}}5,000,000 -\\   10,000,000\end{tabular} & 4.2 & 2.7.5 & 20/08/2017 \\
					11 & Mozilla Stumbler & Tools & 50,000 - 100,000 & 4.5 & 1.8.5 & 20/08/2017 \\
					12 & XOWA & \begin{tabular}[c]{@{}l@{}}Books\\   and reference books\end{tabular} & 1,000 - 5,000 & 3.9 & \begin{tabular}[c]{@{}l@{}}2.1.173-r-\\   2017-06-25\end{tabular} & 21/08/2017 \\
					13 & Journal with Narrate & Lifestyle & 50,000 - 100,000 & 4.2 & 2.4.0 & 08/09/2017 \\
					14 & Wikimedia Commons & Photography & 10,000 - 50,000 & 4.3 & 2.4.2 & 21/08/2017 \\
					15 & WordPress & Social & \begin{tabular}[c]{@{}l@{}}5,000,000 -\\   10,000,000\end{tabular} & 4.2 & 8.0 & 24/08/2017 \\
					16 & GnuCash & Finance & 100,000 - 500,000 & 4.4 & 2.2.0 & 21/08/2017 \\
					17 & My Expenses & Finance & 500,000 - 1,000,000 & 4.4 & 2.7.9 & 24/08/2017 \\
					18 & FBReader & \begin{tabular}[c]{@{}l@{}}Books\\   and reference books\end{tabular} & \begin{tabular}[c]{@{}l@{}}10,000,000 -\\   50,000,000\end{tabular} & 4.5 & 2.8.2 & 24/08/2017 \\
					19 & K-9 Mail & Communication & \begin{tabular}[c]{@{}l@{}}5,000,000 -\\   10,000,000\end{tabular} & 4.2 & 5.207 & 24/08/2017 \\
					20 & MAPS.ME & \begin{tabular}[c]{@{}l@{}}Travel\\   \& Local\end{tabular} & \begin{tabular}[c]{@{}l@{}}10,000,000 -\\   50,000,000\end{tabular} & 4.5 & 7.4.5-Google & 24/08/2017 \\
					21 & Omni Notes & Productivity & 100,000 - 500,000 & 4.4 & 5.3.2 & 24/08/2017 \\
					22 & Hubble Gallery & Education & 50,000 - 100,000 & 4.6 & 1.5.1 & 08/09/2017 \\
					23 & Prey & Tools & 1,000,000 - 5,000,000 & 4.2 & 1.7.7 & 24/08/2017 \\
					24 & Forecastie & Weather & 5,000 - 10,000 & 4.3 & 1.2 & 24/08/2017 \\
					25 & Twidere for Twitter & Social & 100,000 - 500,000 & 4.1 & 3.6.24 & 24/08/2017 \\
					26 & Opengur & Entertainment & 50,000 - 100,000 & 4.4 & 4.7.1 & 24/08/2017 \\
					27 & AnkiDroid & Education & 1,000,000 - 5,000,000 & 4.5 & 2.8.2 & 31/08/2017 \\
					28 & Transportr & \begin{tabular}[c]{@{}l@{}}Maps\\   and navigation\end{tabular} & 5,000 - 10,000 & 4.6 & 1.1.8 & 02/10/2017 \\
					29 & Ouroboros & Communication & 10,000 - 50,000 & 3.5 & 0.10.5.1 & 01/10/2017 \\
					30 & EarthViewer Beta & Personalization & 10,000 - 50,000 & 4.4 & 0.6.1-BETA & 18/09/2017 \\
					31 & Open Weather & Weather & 1,000 - 5,000 & 3.5 & 4.4 & 28/09/2017 \\
					32 & Cannonbal & \begin{tabular}[c]{@{}l@{}}Game\\   - Word\end{tabular} & 1,000 - 5,000 & 4.4 & 1.0.2 & 10/09/2017 \\
					33 & OpenBikeSharing & \begin{tabular}[c]{@{}l@{}}Maps\\   and navigation\end{tabular} & 1,000 - 5,000 & 4.6 & 1.10.0 & 10/09/2017 \\
					34 & c:geo & Entertainment & 1,000,000 - 5,000,000 & 4.4 & 2017.08.23 & 10/09/2017 \\
					35 & PAT Track & \begin{tabular}[c]{@{}l@{}}Maps\\   and navigation\end{tabular} & 10,000 - 50,000 & 4.1 & 7.0.6 & 10/09/2017 \\
					36 & Stepik & Education & 50,000 - 100,000 & 4.8 & 1.42 & 18/09/2017 \\
					37 &RunnerUp & \begin{tabular}[c]{@{}l@{}}Health\\   \& Fitness\end{tabular} & 10,000 - 50,000 & 4 & 1.2 & 10/09/2017 \\
					38 &Wake You in Music & Tools & 10,000 - 50,000 & 3.6 & 1.1.1 & 19/09/2017 \\
					39 &Habitica: Gamify your Tasks & Productivity & 500,000 - 1,000,000 & 4.3 & 1.1.6 & 19/09/2017 \\
					40 &Openshop.io 1.0 & Bussiness & 1,000 - 5,000 & 4 & 1.2 & 19/09/2017 \\
					41 &Glucosio & Medicine & 10,000 - 50,000 & 4.2 & 1.4.0 & 29/09/2017 \\
					42 &Signal Private Messenger & Communication & \begin{tabular}[c]{@{}l@{}}5,000,000 -\\   10,000,000\end{tabular} & 4.6 & 4.11.5 & 02/11/2017 \\
					43 &Tasks: Astrid To-Do List Clone & Productivity & 100,000 - 500,000 & 4.4 & 4.9.14 & 02/10/2017 \\
					44 &Materialistic - Hacker News & \begin{tabular}[c]{@{}l@{}}News\\   \& Magazines\end{tabular} & 50,000 - 100,000 & 4.8 & 3.1 & 26/09/2017 \\
					45 &Kontalk Messenger & Communication & 10,000 - 50,000 & 4.3 & 4.1.0 & 02/11/2017 \\
					46 &Open Food Facts & \begin{tabular}[c]{@{}l@{}}Health\\   \& Fitness\end{tabular} & 100,000 - 500,000 & 4 & 0.7.4 & 28/09/2017 \\
					47 &RedReader & \begin{tabular}[c]{@{}l@{}}News\\   \& Magazines\end{tabular} & 50,000 - 100,000 & 4.6 & 1.9.8.2.1 & 10/10/2017 \\
					48 &Tram Hunter & \begin{tabular}[c]{@{}l@{}}Travel\\   \& Local\end{tabular} & 100,000 - 500,000 & 4.6 & 1.7 & 09/10/2017 \\
					49 & PocketHub for GitHub & Productivity & 10,000 - 50,000 & 3.4 & 0.3.1 & 09/10/2017 \\
					50 & Kickstarter & Social & 1,000,000 - 5,000,000 & 4.5 & 1.6.1 & 28/09/2017 \\
					\bottomrule
			\end{tabular}}
		\end{center}
		
	\end{minipage}
\end{table}

\subsection{Data Collection}
\label{sec:design-manual}
The testing and analysis process  of the selected apps was organized in three-steps: (i) scenarios design, (ii) scenarios execution, and (iii) taxonomy building. The process was carried out by three authors with experience in mobile apps development (hereinafter referred as ``analysts") and took about six months of work with partial dedication. Details about each of the three steps in the data collection are reported in the following.

\subsubsection{Scenarios design} Each application was assigned to one of the three analysts, who had as first task to explore the application and the features it provides. Then, the analyst had to fill-in a template composed by two parts:

\begin{enumerate}[leftmargin=1.5em]
\item \textit{App information:}  Which contains for the given app  (i) its purpose, (ii) its category, (iii) the link to its Google Play page, and (iv) the link to its source code repository. 
\item \textit{Scenarios:} The set of scenarios to be executed on the app during the testing. We define a scenario as \emph{a specific feature to be executed in the app in a specific context}. With ``context'', we refer to the connectivity conditions of the app (\eg \emph{execute the feature when the WiFi connection is off}, \emph{turn off the connection after starting the execution of the feature}, \etc). In summary, a scenario includes the description of the feature to be executed and the steps to follow to reproduce the scenario in the desired context (\eg which connection to turn on/off at a certain moment).
\end{enumerate}

\begin{figure}[t]
	\begin{center}
		\includegraphics[width=0.7\columnwidth]{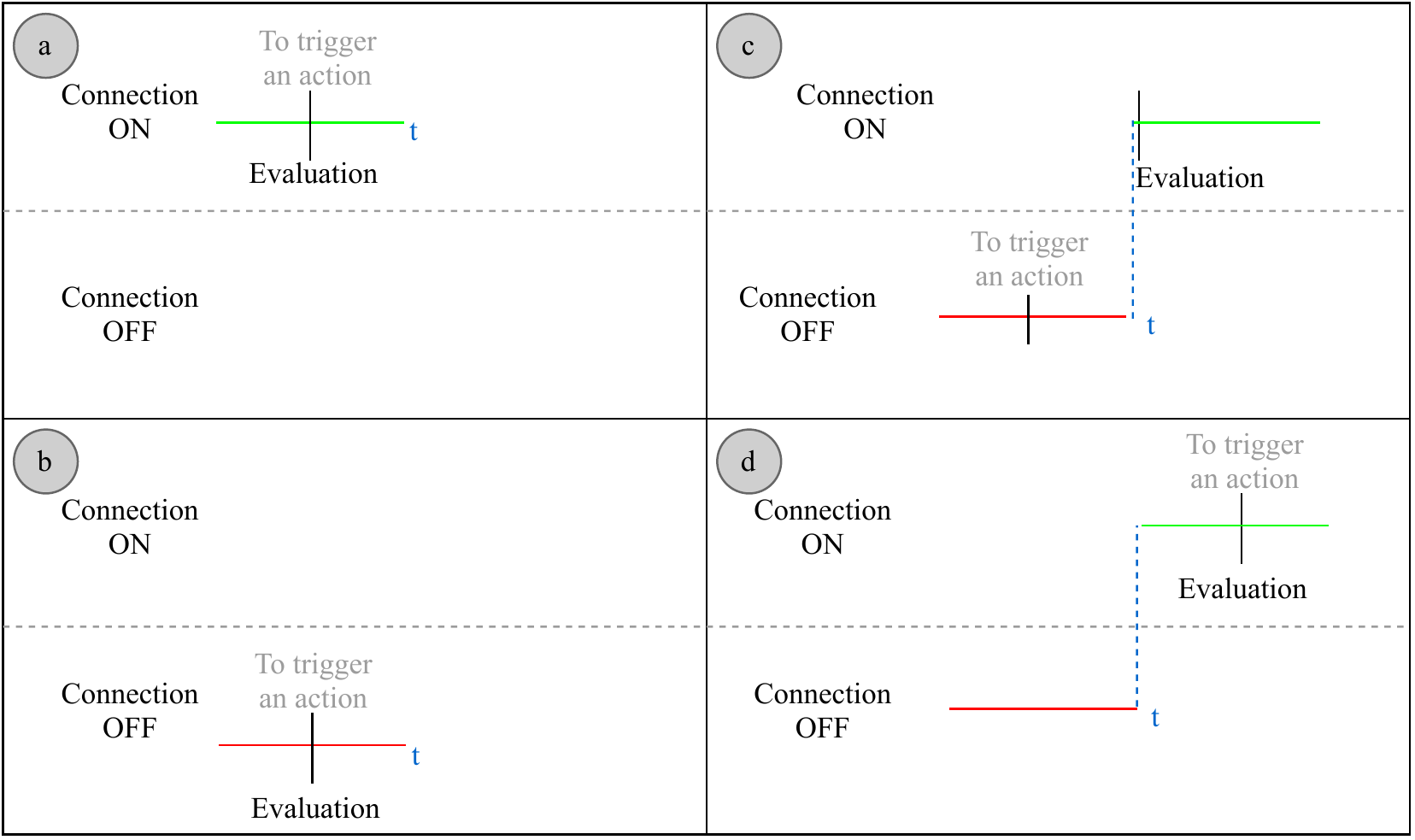}
		\caption{Types of scenarios for testing eventual connectivity.}
		\label{fig:four}
	\end{center}
\vspace{-0.5cm}
\end{figure}

There are six different types of scenarios the  analysts followed for defining the execution steps; the scenarios were identified by letters: \textit{a, b, c, d, e, f}. Four of them, depicted in \figref{fig:four}, were defined for each app. In \figref{fig:four}, the blue line in each scenario represents time and the vertical red line represents the time of an event which can be (i) the execution of the feature, and (ii) the evaluation of the app's behavior. Finally, a vertical dashed line represents a transition between connectivity states (\eg from \emph{connection on} to \emph{connection off}). 

Scenario \textit{a} evaluates a feature using Internet or network connection (connection state) ON. In scenario \textit{b}, the feature was tested with the connection state OFF (\ie Airplane mode on). 
In scenario \textit{c} the goal was to execute the feature while the connection is OFF, and then turning the connection ON after executing the feature. Finally, the scenario \textit{d} evaluates if the app has a normal behavior when executing the feature right after turning ON the connection. These scenarios allow to test cases in which the features that rely on back-end services include the implementation of local caches (in the devices), local queues for later dispatching of messages when the connection is recovered, \etc

There are two additional scenarios (\textit{e} and \textit{f}) that were only defined for specific apps that can support them. Scenario \textit{e} aims at testing features related to SMS sending. For this action, network connection is usually required. In scenario \textit{e}, we trigger the sending process with Internet connection (\ie WiFi) but without phone network connection to test the behavior of the application. For example, the Glucosio app (\textit{v1.4.0}) allows users to invite friends via SMS message. When a user tries to send the invitation without both Network and Internet connection an error message reports ``\textit{Message failed to send}''. If a user has Internet connection but no Network connection, the application shows a success message saying ``\textit{Your invitation has been sent}''. However, the message is not delivered.

Scenario \textit{f} covers app-specific cases that were defined by the analysts. For example, in Hubble Gallery app (\textit{v1.5.1}), when a user tries to see the details of an image the application request a large amount of data. Therefore, one of the \textit{f} scenarios for this app consists of removing the available connection while the download process is in progress.

A total of 971 scenarios were defined for the 50 apps. Note that for one app we can have multiple $x$-type scenarios (where $x$ is one of the six types of scenario we described) involving different features. The minimum number of steps found in a scenario is one, for the apps that present a network-dependent activity  when launched (\eg maps, news feeds), and a maximum of eight steps. These numbers do not take into account the steps required to set the network state to be tested (\eg turn on airplane mode).

\vspace{-0.3cm}
\subsubsection{Scenarios execution} \vspace{-0.3cm} Once the scenarios for an app were designed, an analyst executed each of them at least twice by running the app on a physical device. This re-execution process was performed to avoid issues related to inconsistent behaviours. All the executions were made under the same WiFi connection, and the offline cases were obtained by activating the airplane mode. The mobile devices used for this step were a Samsung S6, an Asus Zenphone 2, and a Motorola Moto G, all equipped with the Android API level 25. During the execution, each analyst collected (i) the specific steps followed to execute a scenario (including screenshots of the app and a video of the execution), (ii) the results obtained when executing the scenario, and (iii) the issues exhibited in the app.

Regarding the issues, those were documented by assigning them a description, a name, and a tag. For instance, the ``Redirection to a different application without connectivity check'' tag was assigned to the cases in which the app does not check the availability of a connection before redirecting the user to a different view or app that requires the connection. This lack of verification derives in views or apps displaying errors (or displaying nothing) because of the assumption of an existent connection. Every-time a new tag was assigned, the other analysts were notified to verify whether the new tag also applied to apps already analyzed. Note that multiple tags could be assigned to the same issue.

\vspace{-0.4cm}
\subsection{Taxonomy building}\vspace{-0.3cm} After executing all the scenarios, we analyzed the assigned tags to build a taxonomy categorizing the issues exhibited in the apps. The taxonomy was defined by the three analysts through multiple open meetings. Once the taxonomy was built, for each scenario exhibiting an issue we analyzed the source code of the corresponding apps to identify the parts involved in the issue and the performed implementation choices. Using this information, we derive a set of lessons learned describing good practices that should be followed by developers to avoid eventual connectivity issues. All the created scenarios, the documents resulting from their execution, and an extensive list of examples for the detected issues are available in our online appendix \cite{onlineAppendix}. 

\vspace{-0.4cm}
\subsection{Impacted Factors}\label{sec:impFac}
\vspace{-0.3cm}

As it was mentioned previously, we enhanced our results by analyzing  the impact of the identified issues over different quality attributes. In order to do this, two authors used the list of quality attributes depicted in the ISO/IEC 25010 standard to evaluate separately the impact of each ECn category using a 4 values scale (\ie High, Medium, Low, None). Once both authors finished this step, they removed the quality attributes that where not impacted by any ECn category and compared the assigned values for the remaining ones. Additionally, they solved the conflicts by presenting examples of issues that support the assigned value. After solving the conflicts the obtained classification was presented to other two authors to validate it. At the end, they identified 7 impacted quality attributes: (i) Functionality, (ii) Performance, (iii) User Experience, (iv) User Interface Aesthetics, (v) Availability, (vi) Resource integrity and Consistence, and (vii) Testability. 
%\vspace{-0.3cm}
\begin{figure}[H]
	\centering
		\includegraphics[angle=90,origin=c,width=\linewidth]{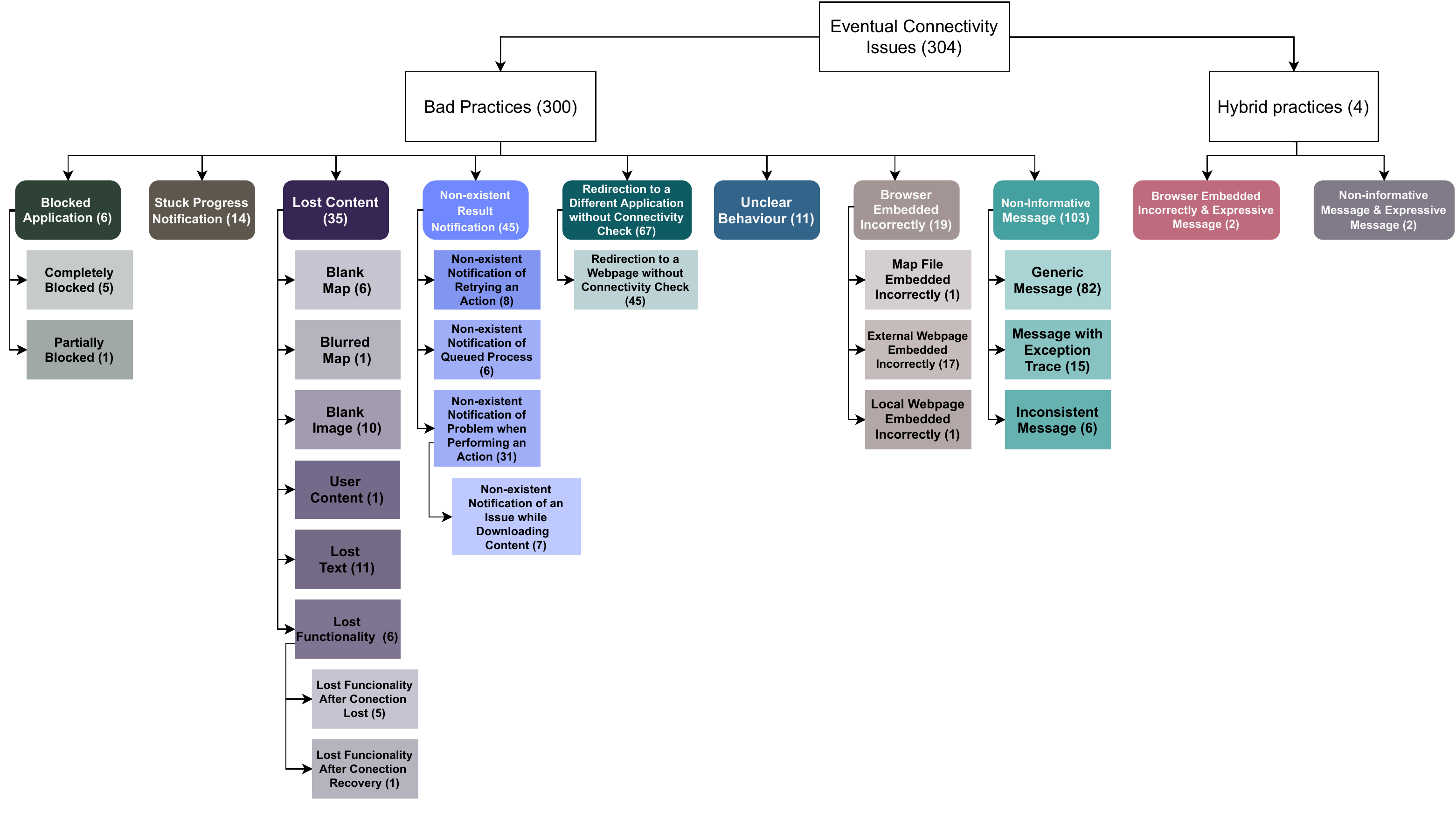}
%		\vspace{-0.5cm}
		\caption{Taxonomy of eventual connectivity  issues in the 50 analyzed apps. Each node contains the tag of the corresponding issue and the number of instances we found for each issue.}
		\label{fig:taxonomy}
\vspace{-0.4cm}
\end{figure}%

\section{Results}
\label{sec:results}

\figref{fig:taxonomy} depicts the taxonomy of eventual connectivity issues, which can be described as errors, usability issues, or confusing behaviors of an app in scenarios where a connection is required. The taxonomy includes high level categories (represented by rounded squares) and low level categories (represented by rectangles). We found 21 low-level types of issues, grouped into 10 high-level categories. Our taxonomy includes a total of 304 issues identified during our analysis\&testing process. Note that the high level categories are organized into two sub-trees. The first, representing the bulk of our taxonomy, is related to bad practices we observed that resulted in connectivity issues. The second, named ``Hybrid practices'', groups cases in which we identified an issue in the tested app but also accompanied by a correct behavior. For example, when a connectivity issue happened, the app shows both a non-informative message \textbf{and} an expressive and useful message. \fref{tab:sumInstIssue} summarizes the number of instances and affected apps for each ECn issue presented in the taxonomy. Additionally, we link the ECn issues with the apps' IDs presented in \fref{tab:one}.

%!TEX root = ../main.tex
\begin{table}[h!]
\caption{Summary of amount of instances and affected apps per ECn issues}
\label{tab:sumInstIssue}
\begin{minipage}{0.5\linewidth}
\centering{
\begin{tabular}{|m{4.5cm} c c m{3cm}|}
\hline
\textbf{ECn Issue}&\textbf{\# Instances}&\textbf{\# Apps}&\textbf{Applications} \\
\hline
\hline
\rowcolor{black} \color{white} Blocked Application & 6 & 4 & \\
\hline
\rowcolor{lightgray} \color{black} Completely Blocked & 5 & 3 & 7, 49, 50\\
\rowcolor{lightgray} Partially Blocked & 1 & 1 & 6\\
\hline
\hline
\rowcolor{black} \color{white} Stuck Progress Notification & 14 & 9 & 2, 10, 23, 31, 32, 36, 39, 46, 49  \\
\hline
\hline
\rowcolor{black} \color{white} Lost Content & 35 & 18 &   \\
\hline
\rowcolor{lightgray} \color{black} Blank Map & 6 & 4 & 6, 33, 34, 35 \\
\rowcolor{lightgray} Blurred Map & 1 & 1 & 37  \\
\rowcolor{lightgray} Blank Image & 10 & 9 & 10, 15, 18, 22, 26, 30, 36, 39, 50 \\
\rowcolor{lightgray} User Content & 1 & 1 & 39  \\
\rowcolor{lightgray} Lost Text & 11 & 7 & 22, 31, 36, 39, 44, 46, 50 \\
\rowcolor{lightgray} Lost Functionality & 6 & 5 &   \\
Lost Functionality after Connection Lost & 5 & 5 & 21, 35, 37, 39, 44 \\
Lost Functionality after Connection Recovery & 1 & 1 & 39  \\
\hline
\hline
\rowcolor{black} \color{white} Non-existent Result Notification & 45 & 26 &  \\
\hline
\rowcolor{lightgray} \color{black} Non-existent Result Notification of Retrying an Action & 8 & 4 & 41, 44, 48, 49 \\
\rowcolor{lightgray} Non-existent Result Notification of Queued Process & 6 & 4 & 4, 14, 19, 44 \\
\rowcolor{lightgray} Non-existent Result Notification of Problem when Performing an Action& 31 & 20 & 8, 15, 21, 25, 26, 30, 31, 35, 36, 37, 38, 39, 45, 46, 50 \\
Non-existent Result Notification of an Issue while Downloading Content & 7 & 5 & 4, 5, 10, 12, 20 \\
\hline
\hline
\rowcolor{black} \color{white} Redirection to a Different Application without Connectivity Check & 67 & 39 & 9, 10, 26, 29, 31, 36  \\
\hline
\rowcolor{lightgray} \color{black} Redirection to a Webpage without Connectivity Check & 45 & 33 & 1, 2, 4, 6, 7, 8, 11, 12, 13, 14, 15, 16, 17, 18, ...\\%18, 19, 20, 21, 22, 23, 24, 30, 32, 33, 34, 35, 37, 38, 39, 41, 42, 43, 45, 47  \\
\hline
\hline
\rowcolor{black} \color{white} Unclear Behaviour & 11 & 7 & 1, 2, 7, 8, 11, 40, 41 \\
\hline
\hline
\rowcolor{black} \color{white} Browser Embedded Incorrectly & 19 & 13 &  \\
\hline
\rowcolor{lightgray} \color{black} Map File Embedded Incorrectly & 1 & 1 & 24 \\
\rowcolor{lightgray} External Webpage Embedded Incorrectly & 17 & 11 & 5, 9, 10, 14, 23, 25, 26, 27, 39, 47, 49 \\
\rowcolor{lightgray} Local Webpage Embedded Incorrectly & 1 & 1 & 18 \\
\hline
\hline
\rowcolor{black} \color{white} Non-Informative Message & 103 & 38 &  \\
\hline
\rowcolor{lightgray} \color{black} Generic Message & 82 & 33 & 1, 2, 3, 4, 5, 6, 8, 14, 15, 16, 18, 19, 22, ...\\%23, 24, 25, 27, 28, 29, 30, 31, 32, 33, 34, 36, 38, 39, 40, 41, 42, 43, 45, 49 \\
\rowcolor{lightgray} Message with Exception Trace & 15 & 8 & 7, 12, 19, 34, 44, 46, 48, 49 \\
\rowcolor{lightgray} Inconsistent Message & 6 & 6 & 6, 7, 14, 23, 34, 38 \\
\hline
\hline
\rowcolor{black} \color{white} Browser Embedded Incorrectly \& Expressive Message & 2 & 1 & 44  \\
\hline
\hline
\rowcolor{black} \color{white} Non-informative Message \& Expressive Message & 2 & 1 & 44  \\
\hline
\end{tabular}}
\end{minipage} 
\end{table}

In order to highlight the more common issues identified within the study, we only discuss the categories in our taxonomy having more than one instance. Additionally, we present qualitative examples of code snippets that generate the identified issue. In our online appendix \cite{onlineAppendix} we provide the complete list of eventual connectivity issues we identified, the corresponding code snippets, and videos exhibiting the issues.

%!TEX root = ../main.tex

\subsection{Bad Practices}

\subsubsection{Blocked Application (BA)}
In this category we classified scenarios where the app gets blocked (or crashes) due to a connectivity problem. This type of issue usually happens when the user tries to perform an action requiring network or Internet connection and the GUI is blocked as result of the lack of connection. The app and/or the device does not respond to the user's commands and it is necessary to restart the app or to wait for a long time until it unblocks. We divided this high-level category into two types: \textit{Completely blocked} and \textit{Partially Blocked}.  

\begin{figure}[h!]
	\centering
	\subfigure[Blocked Application $\rightarrow$ Completely Blocked]
	{
		\label{sfig:bacbexample}
		\includegraphics[width=0.22\textwidth]{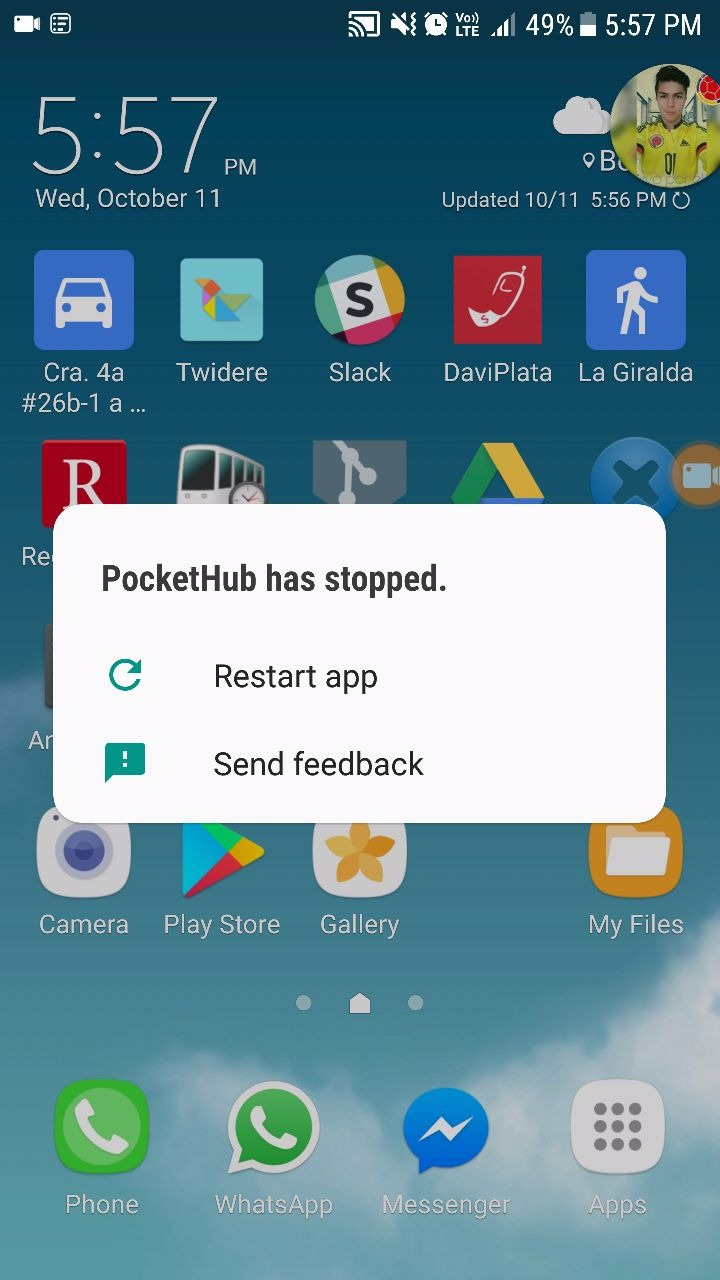}
	}
	\subfigure[Blocked Application $\rightarrow$ Partially Blocked]
	{
		\label{sfig:bapbexample}
		\includegraphics[width=0.22\textwidth]{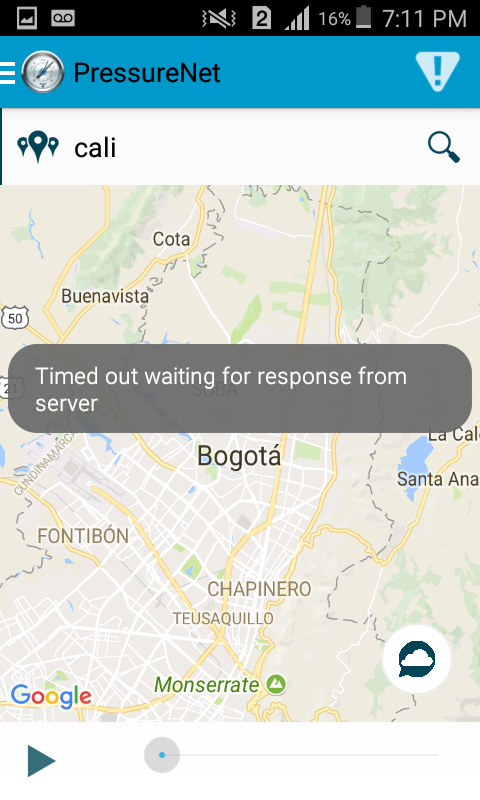}
	}
	\subfigure[Stuck Progress Notification - Pockethub]
	{
		\label{sfig:pbspnexample}
		\includegraphics[width=0.22\textwidth]{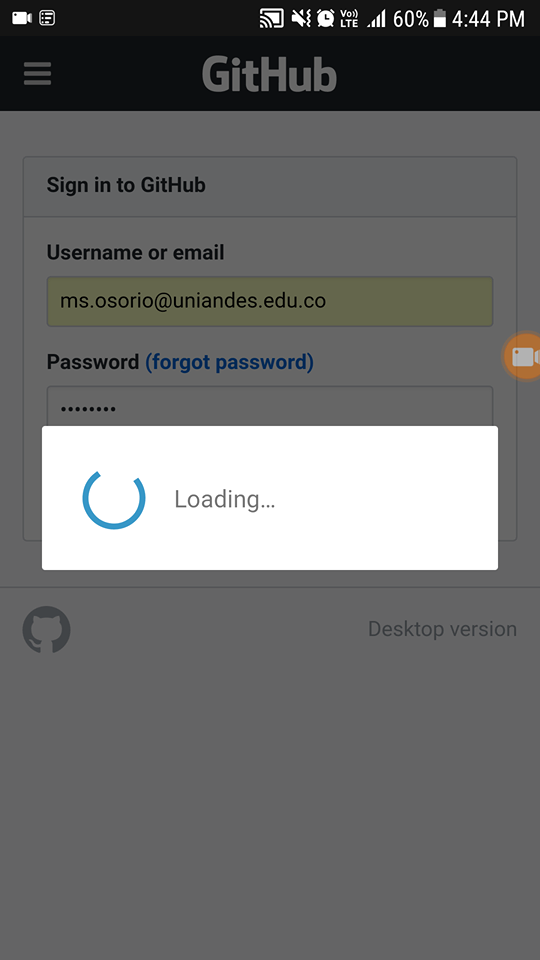}
	}
	\caption
	{
		Examples of connectivity issues regarding Blocked Application and Stuck Progress Notification (1)
	}
	\label{fig:oap1}
\end{figure}
%\vspace{-0.3cm}

 \textit{Completely Blocked (CB)} refers to cases in which it is necessary to restart the app after it gets blocked. This case implies a crash or an Application-Not-Responding error (ANR), which means that the app stops and needs to be launched again. \sfigref{sfig:bacbexample} depicts the bug found in PocketHub \textit{(v0.3.1)}, with the app crashing because it is not properly handling the lack of connectivity. When a user selects the Repositories tab, a new request to a backend service is made. However, there is no code for handling errors in the request/response. To get more details about the code-level issue, we inspected the source code and we found that the app is missing a null check for a returning value. 
In this case (see \listref{lst:bacbStart}, lines 112 and 113), the application uses the \texttt{getRepository}  method for a service generated using Retrofit library. Therefore, due to connectionless state, \texttt{getRepository} returns a null value. Because of this, the concatenated calls starting at line 114 fall into a \texttt{NullPointerException} that is not properly handled. 

\noindent
\begin{minipage}{\linewidth}
	\begin{lstlisting}[language={Java}, label={lst:bacbStart}, caption={Code snippet from PocketHub (app/src/main/java/com/github/pockethub/android/ui/repo/RepositoryViewActivity.java) showing the reasons behind an example of CB.}, firstnumber=106]
if (owner.avatarUrl() != null && RepositoryUtils.isComplete(repository)) {
	checkReadme();
} else {
	avatars.bind(getSupportActionBar(), owner);
	loadingBar.setVisibility(View.VISIBLE);
	setGone(true);
	ServiceGenerator.createService(this, RepositoryService.class)
		.getRepository(repository.owner().login(), repository.name())
		.subscribeOn(Schedulers.io())
		.observeOn(AndroidSchedulers.mainThread())
		.compose(this.bindToLifecycle())
		.subscribe(response -> {
			repository = response.body();
			checkReadme();
		}, e -> {
			ToastUtils.show(this, R.string.error_repo_load);
			loadingBar.setVisibility(View.GONE);
		});
}
	\end{lstlisting}
\end{minipage}\\

The \textit{Partially Blocked (PB)} type covers the cases in which the app does not respond to commands given by the user, but just for a short period of time. After waiting, the app just keeps working normally and sometimes displays a message indicating that there was an issue. This happened in the PressureNet app (see \sfigref{sfig:bapbexample}).

\subsubsection{Stuck Progress Notification (SPN)}
The \textit{Stuck Progress Notification} type is exhibited when a progress notification gets stuck in the GUI (not necessarily blocking the app) because of a connection problem. For example, this happens in PocketHub \textit{(v0.3.1)} (see \sfigref{sfig:pbspnexample}): The app shows a loading dialog while singing in and there is no connectivity. However, the dialog gets stuck. PocketHub uses an embedded \texttt{WebView} to show the GitHub login page. In PocketHub the \texttt{Web\-View\-Client.\-on\-Page\-Started} method is overridden to show the loading dialog every time a webpage is requested (see \listref{lst:phbspnStart}); and the \texttt{WebViewClient.onPageFinished} method is overridden to dismiss the dialog when a requested page URL finishes loading.  

The \texttt{onReceiveError} callback is supposed to define the corresponding action to execute. However, PocketHub does not override the method which is the right place for dismissing the loading dialog when there is a connectivity error.

\begin{lstlisting}[language={Java}, label={lst:phbspnStart}, caption={Code snippet from PocketHub showing the reasons for an example of SPN.}, firstnumber=59]
@Override
public void onPageStarted(android.webkit.WebView view, String url, Bitmap favicon) {
dialog.show();
}
...
@Override
public void onPageFinished(android.webkit.WebView view, String url) {
dialog.dismiss();
}
\end{lstlisting}

\subsubsection{Lost Content (LC)}

\begin{figure}
	\centering
	\subfigure[Lost Content $\rightarrow$ Blank Map]
	{
		\label{sfig:lcbmexample}
		\includegraphics[width=0.22\linewidth]{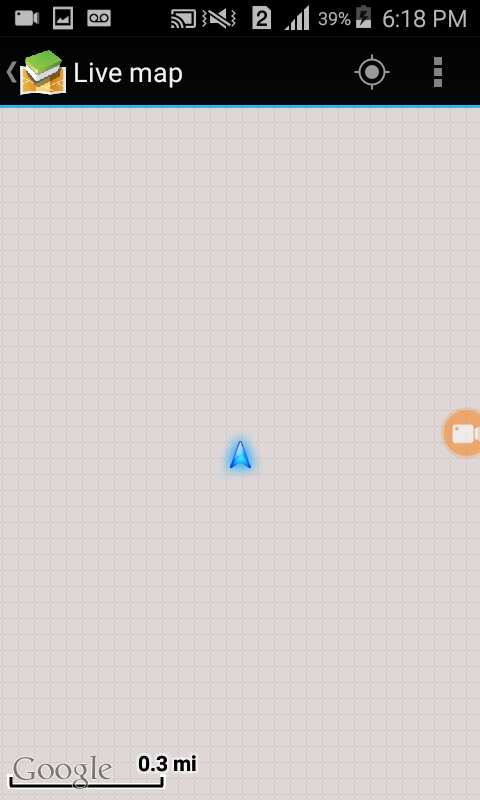}
	}
	\subfigure[Lost Content $\rightarrow$ Blank Image]
	{
		\label{sfig:lcbiexample}
		\includegraphics[width=0.22\linewidth]{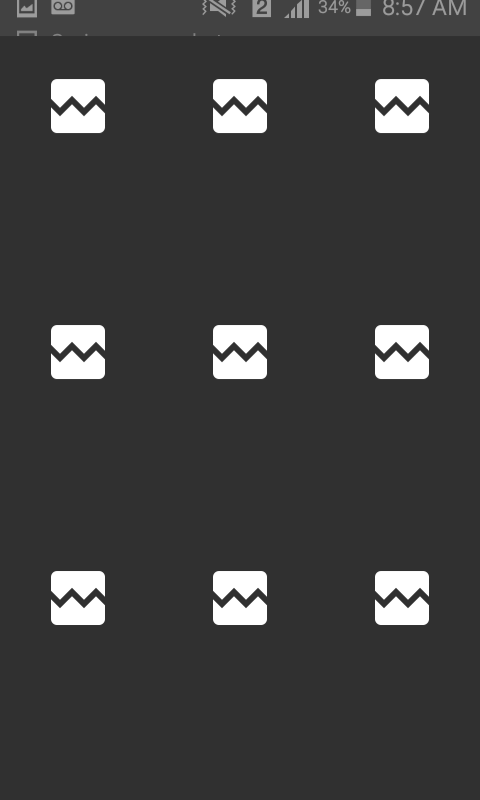}
	}
	\subfigure[Lost Content $\rightarrow$ Lost Text ]
	{
		\label{sfig:lcltexample}
		\includegraphics[width=0.22\linewidth]{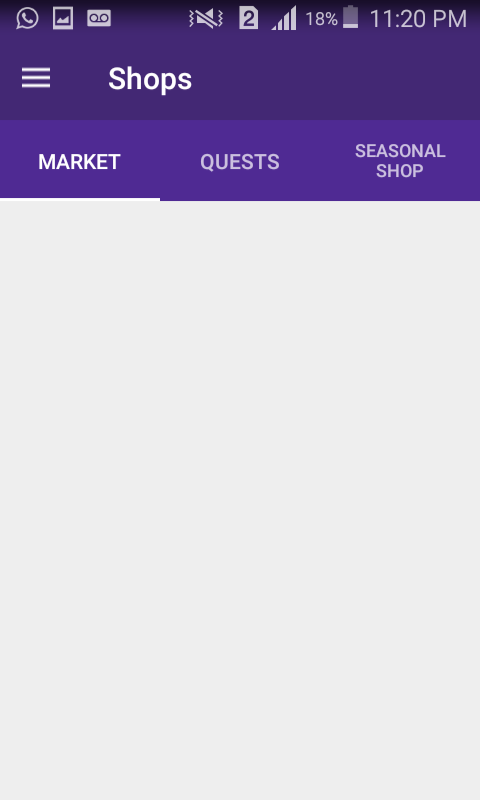}
	}
	\subfigure[Lost Content $\rightarrow$ Lost Functionality]
	{
		\label{sfig:lclfcexample}
		\includegraphics[width=0.22\linewidth]{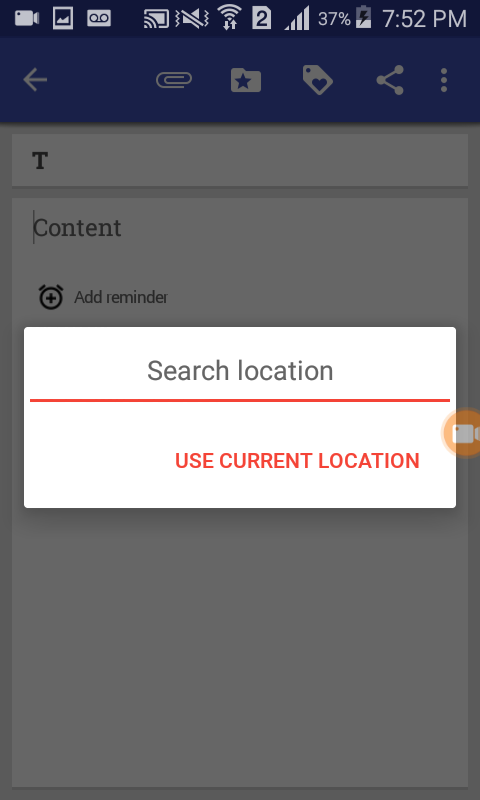}
	}
	\caption
	{
		Examples of connectivity issues in the analyzed apps (3) 
	}
	\label{fig:oap3}
\end{figure}

A \textit{Lost Content} issue happens when an app (i) does not have connectivity, (ii) it shows empty, incomplete, or blurred content where it is supposed to be, and (iii) it does not notify the user about the connectivity problem.

The  \textit{Blank Map (BM)}  type  refers to cases in which an app provides a map functionality, and due to lack of connection, the map is blank (\ie does not show any layer of the map) and there is no message notifying the issue. For instance, the C:Geo app (see \sfigref{sfig:lcbmexample}) allows users to see a ``Live Map''. However, when a user selects this option when connectivity lacks, the app shows a blank space in the view dedicated to the map. After reviewing the code (see \listref{lst:lcbmCGeo}) we found that the map is created and loaded without checking the connection state.
\noindent
\begin{minipage}[t]{\linewidth}
	\begin{lstlisting}[language={Java}, label={lst:lcbmCGeo}, caption={Code snippet that generates a blank map in C:Geo}, firstnumber=433]
// initialize map
mapView = (MapViewImpl) activity.findViewById(mapProvider.getMapViewId());
mapView.setMapSource();
mapView.setBuiltInZoomControls(true);
mapView.displayZoomControls(true);
mapView.preLoad();
mapView.setOnDragListener(new MapDragListener(this));
	\end{lstlisting}
\end{minipage}\\

The \textit{Blurred Map (BMA)} type describes scenarios in which the app still shows a map even in the absence of connection. However, the map is blurred. This behavior was found only once in RunnerUp \textit{(v1.2)}.

The \textit{Blank Image (BI)} low-level type covers the cases in which the app uses external images inside the application and, due to connectionless state, the app displays a blank image (or empty content) where the image is supposed to be.  
For example,  OpenGur\textit{(v4.7.1)} shows a grid with images by using the Android-\-Universal-\-Image-\-Loader library. However, if the app is launched without Internet no image is loaded in the grid (\sfigref{sfig:lcbiexample})  and the user is not notified about the issue. After inspecting the code we found that the issue is generated because (i) the connection state is not validated before filling the grid, and (ii) the developer is not using the caching feature provided by the library (\listref{lst:lcbiOpengur}).

\noindent
\begin{minipage}{\linewidth}
	\begin{lstlisting}[language={Java}, label={lst:lcbiOpengur}, caption={Code snippet for displaying an image using the Android-Universal-Image-Loader library (Opengur)}, firstnumber=48]
protected void displayImage(ImageView imageView, String url) {
	if (imageLoader == null) {
		throw new IllegalStateException("Image Loader has not been created");
	}

	imageLoader.cancelDisplayTask(imageView);
	imageLoader.displayImage(url, imageView, getDisplayOptions());
}
	\end{lstlisting}
\end{minipage}

The \textit{User Content (UC)} type describes scenarios in which, due to lack of connectivity, content generated by the user is lost. This behavior was found only in the chat feature of Habitica \textit{(v1.1.6)}.

The \textit{Lost Text (LT)} low-level type covers the cases in which a text section is not available due to connectionless state. For instance, Habitica app \textit{(v1.1.6)}: when a user tries to enter to the ``Shops'' option without connection (see \sfigref{sfig:lcltexample}), the app shows an empty view. Specifically, Habitica uses the Retrofit library when retrieving the `Shops' option information. As it can be seen in \listref{lst:lcltHabiticapre}, when there is an error it is handled using the ``handleEmptyError" method (see line 132). 

\noindent
\begin{minipage}{\linewidth}
	\begin{lstlisting}[language={Java}, label={lst:lcltHabiticapre}, caption={Code snippet showing the 'Shops' information retrieving process in Habitica}, firstnumber=114]
this.inventoryRepository.retrieveShopInventory(shopUrl)
    .map { shop1 ->
        if (shop1.identifier == Shop.MARKET) {
            val user = user
            if (user != null && user.isValid && user.purchased.plan.isActive) {
                val specialCategory = ShopCategory()
                specialCategory.text = getString(R.string.special)
                val item = ShopItem.makeGemItem(context?.resources)
                item.limitedNumberLeft = user.purchased.plan.numberOfGemsLeft()
                specialCategory.items.add(item)
                shop1.categories.add(specialCategory)
            }
        }
        shop1
    }
    .subscribe(Action1 {
        this.shop = it
        this.adapter?.setShop(it, configManager.shopSpriteSuffix())
    }, RxErrorHandler.handleEmptyError())
	\end{lstlisting}
\end{minipage}\\

Nevertheless, as it can be seen in \listref{lst:lcltHabiticapos}, ``handleEmptyError'' sends a message to the Crashlytics log but does not notify the user.

\noindent
\begin{minipage}{\linewidth}
	\begin{lstlisting}[language={Java}, label={lst:lcltHabiticapos}, caption={Code snippet showing how errors are handled in Habitica}, firstnumber=27]
public static Action1<Throwable> handleEmptyError() {
    //Can't be turned into a lambda, because it then doesn't work for some reason.
    return new Action1<Throwable>() {
        @Override
        public void call(Throwable throwable) {
            RxErrorHandler.reportError(throwable);
        }
    };
}
	\end{lstlisting}
\end{minipage}\\

The \textit{Lost Functionality (LF)} type describes scenarios in which a functionality that requires Internet is missed, or hidden when there is no connectivity, and the user is not notified about the case. Within this category we defined two sub-categories, namely \textit{Lost Functionality After Connection Lost (LFACL)} and \textit{Lost Functionality After Connection Recovery (LFACR).}

The \textit{Lost Functionality After Connection Lost (LFACL)} category describes the behavior in which a feature of an app is no longer functional when connection status is OFF. 
One example from Omni Notes \textit{(v5.3.2)} is depicted in \sfigref{sfig:lclfcexample}. When a user selects the location option  and there is connection, then a dialog is displayed; however, if the same feature is invoked without connectivity, the dialog does not show up and there is no notification about the connection error. In this specific case, the application uses the latest location the phone registered before losing connection. 

\textit{Lost Functionality After Connection Recovery (LFACR)} covers the issues appearing in mobile apps when the following events occur: (i) a user performs an action in the app (with connectivity) and everything works as expected; (ii) the user performs the same action again but this time without connectivity; (iii) the action is not performed because it relies on having a connection (and it is the expected behavior); (iv) the user turns on the required connection (or the connection is back) to perform again the action; (iv) the functionality is not available/functional anymore.
Note that this is similar to \textit{LFACL}. However, while \textit{LFACL} happens when there is no connectivity, in the case of \textit{LFACR} the functionality is ``lost" after recovering connectivity. We found only one instance of this issue in the Habitica app \textit{(v1.1.6)}, that is detailed in our online appendix \cite{onlineAppendix}.

\begin{figure}[h!]
	\centering
	\subfigure[NRN $\rightarrow$ Non-existent notification of retrying an action]
	{
		\label{sfig:nrnnnraexample}
		\includegraphics[width=0.22\linewidth]{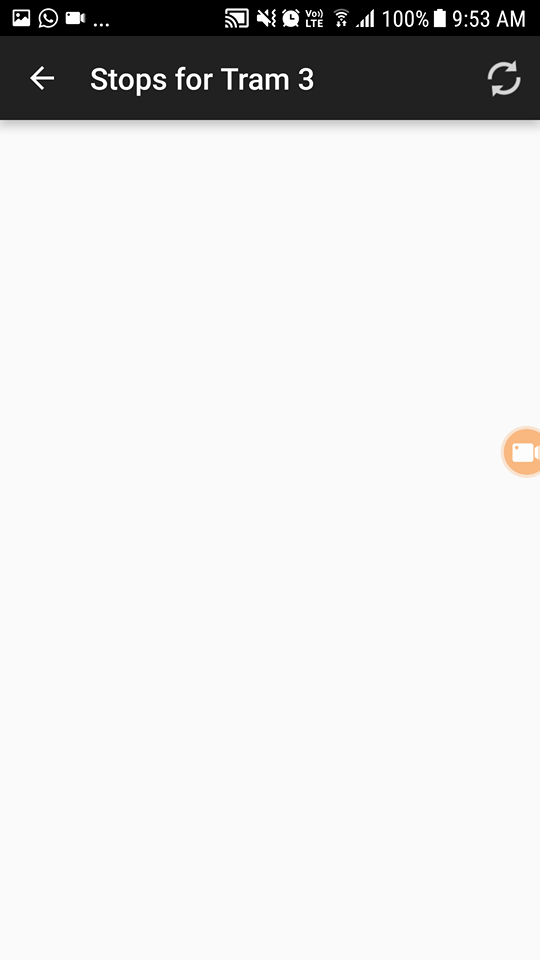}
	}
	\subfigure[NRN $\rightarrow$ Non-existent notification of queued process]
	{
		\label{sfig:urbpexample}
		\includegraphics[width=0.22\linewidth]{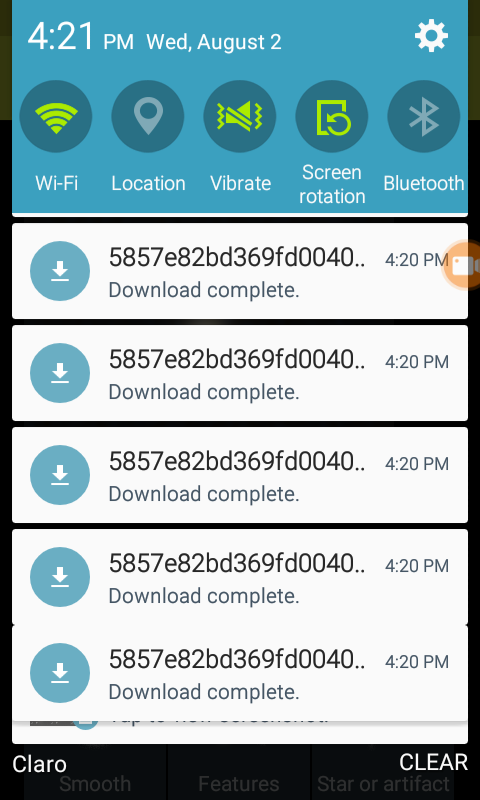}
	}
	\subfigure[NRN $\rightarrow$ Nonexistent notification of problem when performing an action]
	{
		\label{sfig:nrnnnppa}
		\includegraphics[width=0.22\linewidth]{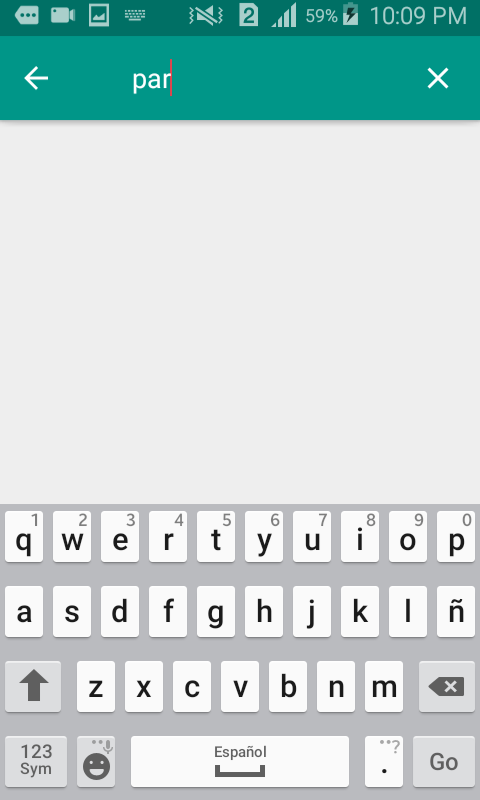}
	}
	\subfigure[NRN $\rightarrow$ Non-existent notification of an issue while downloading content]
	{
		\label{sfig:nrnnndpexample}
		\includegraphics[width=0.22\linewidth]{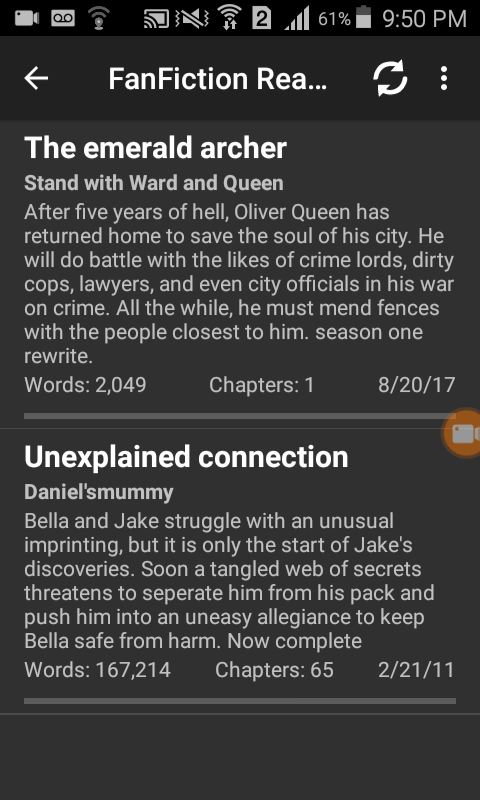}
	}
	\caption
	{
			Examples of connectivity issues in the analyzed apps (4) 
	}
	\label{fig:oap4}
\end{figure}
\subsubsection{Non-existent Result Notification (NRN)} 
In this case the app does not show any result or message indicating that (i) the action executed  by the user or by the system was performed successfully (or not), or (ii) the user must execute a retry/refresh action. We found 45 examples of \textit{NRN} in 26 apps.

The \textit{Non-existent notification of retrying an action (NNRA)} type covers the cases in which a user must retry an action due to connection state but she is not notified about it. TramHunter \textit{(v1.7)} invokes an external service each time a ``Tram Station" is selected to show buses/trams schedule. However, as shown in \sfigref{sfig:nrnnnraexample}, if there is no connection the application neither shows any content nor a message warning that the action must be retried later because there is no Internet connection.

A \textit{Non-existent notification of queued process (NNQP)} issue happens when the following sequence of events occur: (i) the user attempts to perform an action without connection and the app does not execute the background process; then (ii) the connection is recovered and the application executes the actions implemented by the background process. However, the user is not notified about the execution. Galaxy Zoo app \textit{(v1.69)} exhibits this bad practice (see \listref{lst:urbpGalaxyZoo}) when a user attempts to download a galaxy image and there is no connection. In this scenario, the user clicks the ``Download" option, and the download is not executed because of the lack of connection. However, the action is queued for later execution without notifying the user that it will be executed when a connection is established,  and then it is started automatically  without notifying the user (again). Therefore, if the user clicks $n$ times, when the connection is recovered $n$ download requests will be executed. This is an issue that can have implications on resource consumption and usability, because background processes can consume a lot of device resources and the user is not notified about them.
\begin{minipage}{\linewidth}
	\begin{lstlisting}[language={Java}, label={lst:urbpGalaxyZoo}, caption={Code snippet download request handling from the GalaxyZoo app}, firstnumber=195]
final DownloadManager.Request request = new DownloadManager.Request(uri);
request.setNotificationVisibility(DownloadManager.Request.VISIBILITY_VISIBLE_NOTIFY_COMPLETED);

final Activity activity = getActivity();
if (activity == null) {
    Log.error("doDownloadImage(): activity was null.");
    return;
}

final Object systemService = activity.getSystemService(Context.DOWNLOAD_SERVICE);
if (systemService == null || !(systemService instanceof DownloadManager)) {
    Log.error("doDownloadImage(): Could not get DOWNLOAD_SERVICE.");
    return;
}

final DownloadManager downloadManager = (DownloadManager)systemService;
downloadManager.enqueue(request);
	\end{lstlisting}
\end{minipage}
%}

\textit{Non-existent notification of problem when performing an action (NNPPA)} refers to  scenarios in which an app had a problem while performing a connectivity-related action and it does not inform users about the issue. For example, with the Good Weather app \textit{(v4.4)} users can query weather conditions for a specific location. However, when they try to search for a location without Internet, the application is unable to invoke external services for locations that match the queries and it shows an empty list of locations as illustrated in \sfigref{sfig:nrnnnppa}. It is worth noticing that the issue identified by this tag is the lack of notification regarding the disabled internet connection, not the lack of text from the response. Namely, this category should not be confused with the \textit{Lost Text category}.

The \textit{Non-existent notification of an issue while downloading content (NNDC)} issue describes scenarios in which an app provides a download option and there is no notification when an error occurs. This is an important category in our taxonomy since it represents 22.5\% of the issues identified within NNPPA. For example, the FanFiction Reader app \textit{(v1.51)} allows users to check manually if there are updates for the books downloaded before. However, if the user refreshes the book without connectivity,  the application tries to check for new chapters but it fails and then it does not notify the user about the failure, which could make the user thinks the last version is already downloaded. This behavior is depicted in \sfigref{sfig:nrnnndpexample}.

\begin{figure}[h]
		\centering
	\subfigure[RDAC]
	{
		\label{sfig:rwccrdacexample}
		\includegraphics[width=0.22\linewidth]{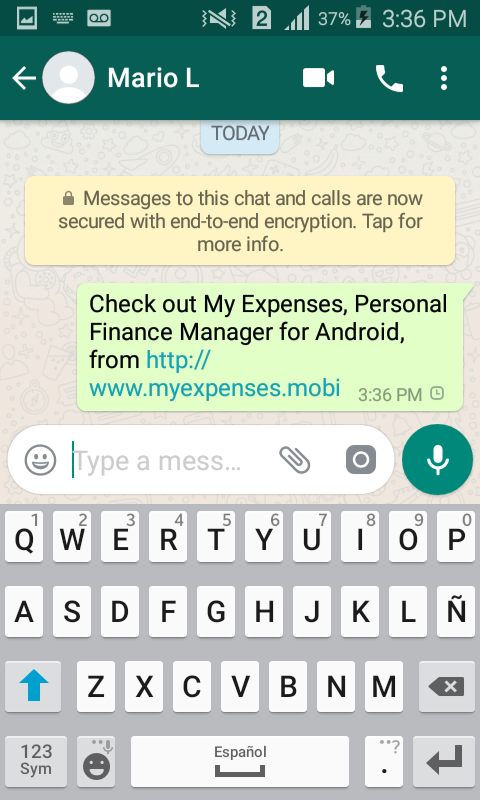}
	}
	\subfigure[RDAC $\rightarrow$ RWPC]
	{
		\label{sfig:rwccrwpcexample}
		\includegraphics[width=0.22\linewidth]{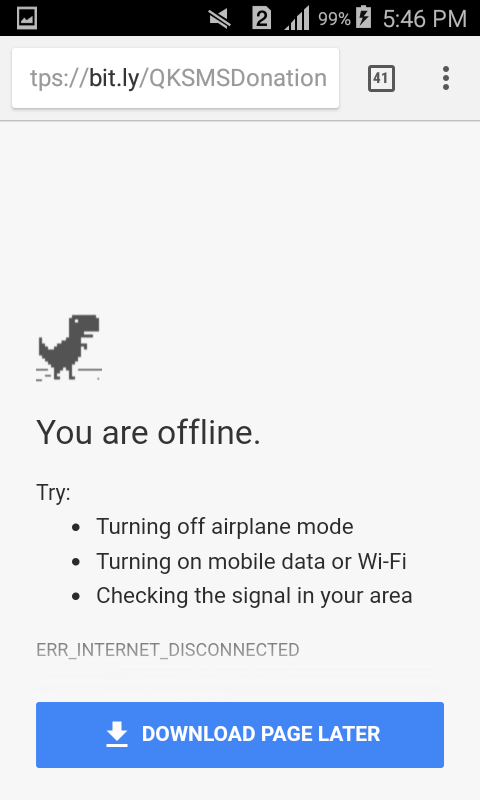}
	}
	\subfigure[UB]{
		\label{sfig:ucbuesexample}
		\includegraphics[width=0.22\linewidth]{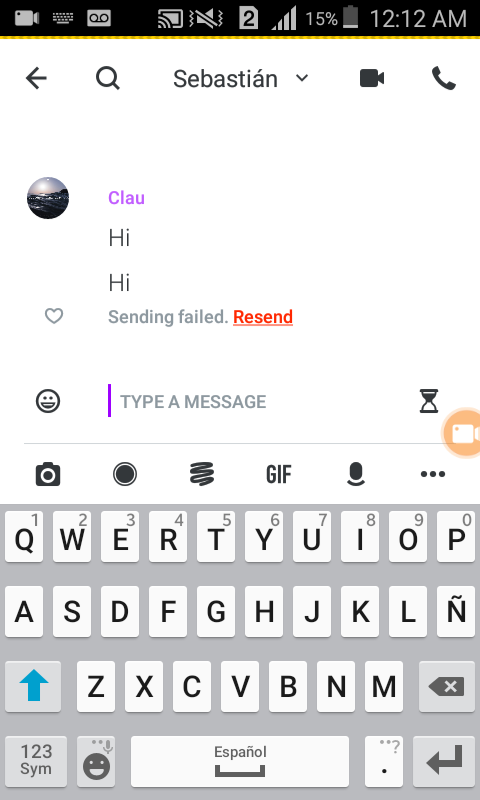}	
	}
	\caption
	{
			Examples of Redirection to a Different Application without Connectivity Check and Unclear Behavior connectivity issues in the analyzed apps 
	}
	\label{fig:oap5}
\end{figure}

\subsubsection{Redirection to a Different Application without Connectivity Check (RDAC)}
This category is the second most frequent in our study. It occurs when an app does not check the connection before redirecting a user to a different view in which Internet connection is required. It is worth noticing that the interaction with the app the user is being redirected to is not necessarily synchronic, nevertheless, the lack of network state validation could impact the usability of the original app. For this category, we identified a special subtype in which the application to which the user is redirected to is an external browser.

In the My Expenses app \textit{(v2.7.9)}, users can ``Tell a friend'' about their My Expenses usage. This feature redirects to the messaging apps in the phone (\eg WhatsApp) without checking connection state. Most of the messaging apps (chats) require connection. Thus, if there is no connection, the responsibility of telling the user that there is no connection relies on the communication apps. From the usability perspective this is not ideal because the app starting the redirection should inform the user about the connection state. \sfigref{sfig:rwccrdacexample} shows an example of RDAC in which the flow is redirected from MyExpenses (without connectivity check) to WhatsApp (note the loading icon in the message because of the connectionless state). As shown in \listref{lst:rwccrdacMyExpenses}, the issue happens since the app sends the \textit{implicit intent} without checking connection state.

\noindent
\begin{minipage}{\linewidth}
	\begin{lstlisting}[language={Java}, label={lst:rwccrdacMyExpenses}, caption={Example of redirection to an app without connectivity check  at MyExpenses}, firstnumber=603]
case R.id.SHARE_COMMAND:
    i = new Intent();
    i.setAction(Intent.ACTION_SEND);
    i.putExtra(Intent.EXTRA_TEXT, Utils.getTellAFriendMessage(this));
    i.setType("text/plain");
    startActivity(Intent.createChooser(i, getResources().getText(R.string.menu_share)));
    return true;
	\end{lstlisting}
\end{minipage}\\
	
A \textit{Redirection to a Web Page without Connectivity Check (RWPC)} describes scenarios in which an app needs to redirect users to the browser to open a web page. An example of this case is in QKSMS \textit{(v2.7.3)}. This app lets users send and receive SMS/MMS. When a user tries to donate with PayPal through QKSMS, it redirects the user to a web page in a browser. If there is no connection, after a while waiting for a response, the browser displays a message declaring that there is no connectivity. \sfigref{sfig:rwccrwpcexample} shows an example of this behavior in QKSMS when redirecting the user to a browser.
This behavior is caused by not validating the connection state before doing the redirection as it can be seen in \listref{lst:rwccrwpcQKSMS}.

\noindent
\begin{minipage}{\linewidth}
	\begin{lstlisting}[language={Java}, label={lst:rwccrwpcQKSMS}, caption={Example of redirection to a web page without connectivity check  at QKSMS}, firstnumber=216]
public void donatePaypal() {
    Intent browserIntent = new Intent(Intent.ACTION_VIEW, Uri.parse("https://bit.ly/QKSMSDonation"));
    mContext.startActivity(browserIntent);
}
	\end{lstlisting}
\end{minipage}\\

\subsubsection{Unclear Behavior (UCB)}
This category refers to the cases in which an app can have different or unexpected behaviors as a consequence of connectivity issues and the reasons are not clear (from the user point of view).

UCBs are exhibited in a situation in which the application, under the same scenario, has different results/responses after performing the same action multiple times. In Wire (a messenger app), when a user tries to send multiple messages to a contact without connectivity, the app shows different behaviors like (see \sfigref{sfig:ucbuesexample}). First, a message of error with a ``Resend'' option is displayed. After sending a few messages, a message saying ``Sending...'' is displayed. This situation confuses the user because the app seems to have different behaviors depending on the number of times that an action is performed.

\subsubsection{Browser Embedded Incorrectly (BEI)}
This category represents the scenarios in which the app redirects the user to an activity with an embedded content and it does not work because there is no connection. We found three sub-types for this category: \textit{Local webpage Embedded Incorrectly}, \textit{External Webpage Embedded Incorrectly}, and \textit{Map File Embedded Incorrectly}.

\begin{figure}[h]
	\centering
	\subfigure[BEI $\rightarrow$ Map File embedded incorectly]
	{
		\label{sfig:beimfeiexample}
		\includegraphics[width=0.22\linewidth]{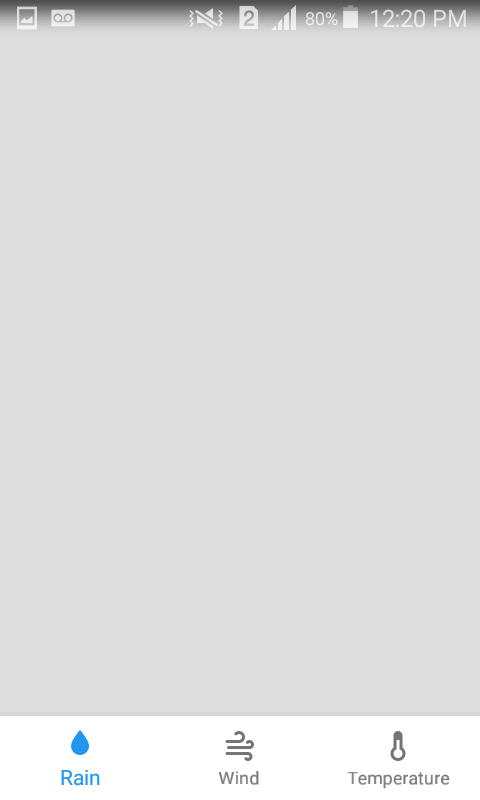}
	}
	\subfigure[BEI $\rightarrow$ External webpage embedded incorrectly]
	{
		\label{sfig:beieweiexample}
		\includegraphics[width=0.22\linewidth]{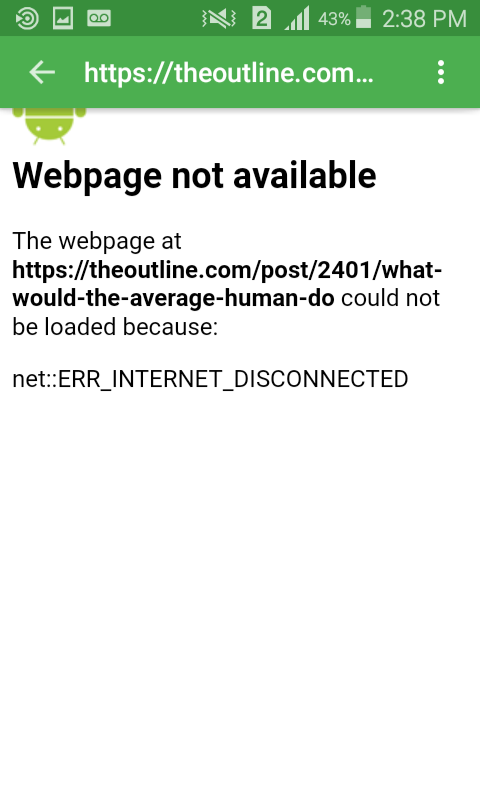}
	}
	\subfigure[BEI $\rightarrow$ Local webpage embedded incorrectly]
	{
		\label{sfig:beilweiexample}
		\includegraphics[width=0.22\linewidth]{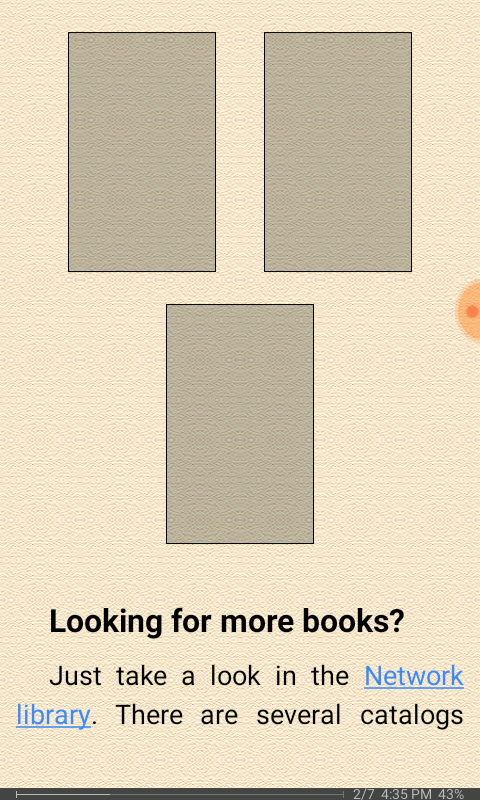}
	}
	\caption
	{
		Occurrences of Anti-Patterns in Studied Applications (7) 
	}
	\label{fig:oap7}
\end{figure}

The \textit{Map File Embedded Incorrectly (MFEI)} type happens when an app opens an Android activity with a local map (\ie there is a local file with the map information) and it does not work because there is no connectivity. An example of this issue was detected in  Forecastie \textit{(v1.2)}, an app for checking surrounding weather conditions (see \sfigref{sfig:beimfeiexample}). It is worth highlighting that unlike \textit{Blank Map} and \textit{Blurred Map}, MFEI uses a local file to display the map, therefore, the ECn issue is generated by the developers.
	
The \textit{External Webpage Embedded Incorrectly (EWEI)} type is exhibited in scenarios in which the app opens an Android Activity with an embedded WebView that points to an external source (URL). RedReader \textit{(v1.9.8.2.1)}, an unofficial open source client for Reddit, is affected by this issue. When a user tries to access an article without connection, the app shows an embedded page with an error stating ``Web page not available. The webpage could not be loaded because: net::ERR\_INTERNET\_DISCONNECTED'' as in \sfigref{sfig:beieweiexample}. This scenario is problematic because (i) the error message shown by the WebView is not user friendly; and (ii) the user has to wait for a time-out until the notice. It is worth noticing that there is a difference between EWEI and RWPC, since EWEI issues do not redirect users to an external app.

The \textit{Local Webpage Embedded Incorrectly (LWEI)} type characterizes the scenarios in which the app opens an Android Activity that has an embedded WebView, but the web content is not displayed because of lack of connectivity. We detected an instance of this type in  FBReader \textit{(v2.8.2)}, a free ebook reader (see \sfigref{sfig:beilweiexample}).

\begin{figure}[h]
	\centering
\subfigure[Non-Informative Message $\rightarrow$ Generic Message]
{
	\label{sfig:nimgmexample}
	\includegraphics[width=0.22\linewidth]{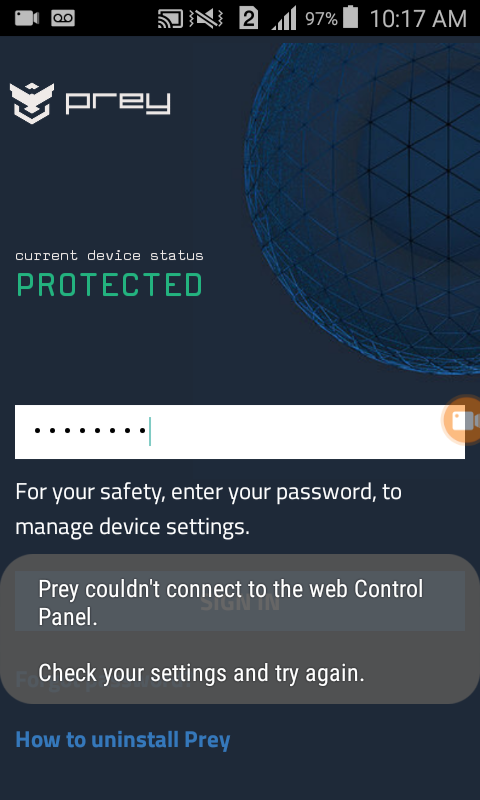}
}
\subfigure[Non-Informative Message $\rightarrow$ Message with Exception Trace]
{
	\label{sfig:nimmetexample}
	\includegraphics[width=0.22\linewidth]{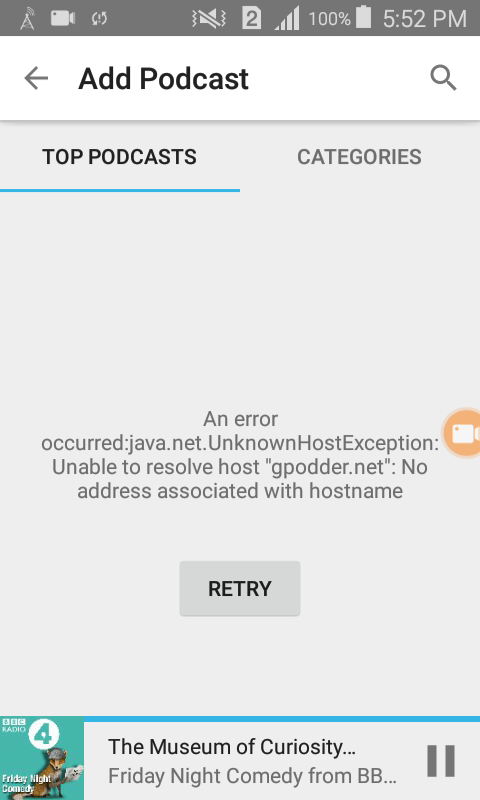}
}
\subfigure[Non-Informative Message $\rightarrow$ Inconsistent Message]
{
	\label{sfig:nimimexampleOpenShop}
	\includegraphics[width=0.22\linewidth]{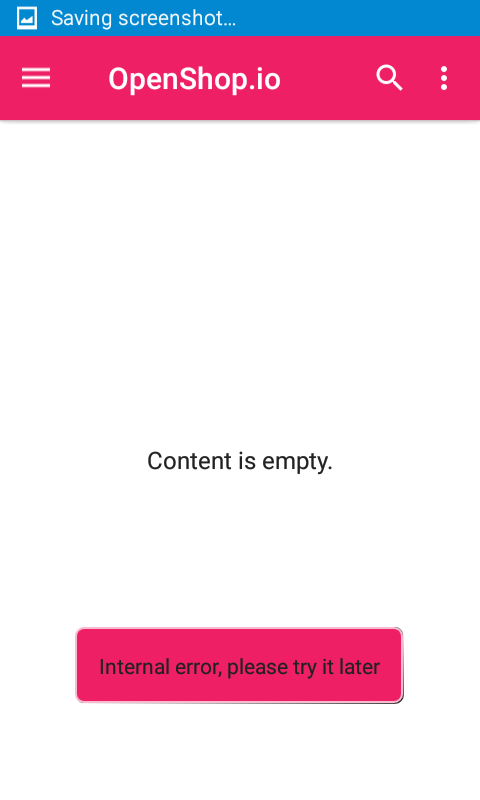}
}
\caption
{
	Examples of connectivity issues (Non-Informative Messages)
}
\label{fig:oap2}
\end{figure}

\subsubsection{Non-Informative Message (NIM)}
A \textit{Non-Informative Message} manifests in apps showing generic, unclear or inconsistent messages when there is a connection problem. Examples of NIMs are  ``An error occurred'' or ``<Exception>: <Exception\_trace>''. A more detailed list of NIMs can be found in our online appendix \cite{onlineAppendix}. We distinguish three low-level types in this category: \textit{Generic Message}, \textit{Message with Exception Trace} and \textit{Inconsistent Message}

The \textit{Generic Message (GM)} type describes those scenarios in which a user is performing an action in the app, a connectivity problem arises and a message is presented asserting that something happened. However, the message does not tell anything about what the problem is, how to solve it, or what is going to happen with the user action. 
In \sfigref{sfig:nimgmexample} we present an example of this behavior at the Prey app \textit{(v1.7.7)}. Since the error handling mechanism in the app is generic (\ie it does not distinguish different types of error), Prey does not have a catch block for specific exceptions, showing a generic message to the user.  Therefore, as it can be seen in \listref{lst:iemgmPrey}, the app throws a general exception called \texttt{PreyException} with the string \texttt{error\_communication\_exception}, that contains the message shown in \sfigref{sfig:nimgmexample}: ``Prey couldn't connect to the web Control Panel. Check your settings and try again". A similar programming error (\ie generic error handling) was found in the Surespot Encrypted Messenger app \textit{(v70)} when using the feature to invite a friend.  Any connection error when invoking the REST URL used in that feature is handled by the \texttt{onFailure} method which always displays ``Could not invite friend, please try again later". 

\noindent
\begin{minipage}{\linewidth}
	\begin{lstlisting}[language={Java}, label={lst:iemgmPrey}, caption={Example of error handling code with generic message  (Prey)}, firstnumber=304]
try {
    String uri=PreyConfig.getPreyConfig(ctx).getPreyUrl().concat("profile.xml");
    PreyHttpResponse response = PreyRestHttpClient.getInstance(ctx).get(uri, parameters, apikey, password);
    xml=response.getResponseAsString();
} catch (Exception e) {
    throw new PreyException(ctx.getText(R.string.error_communication_exception).toString(), e);
}
	\end{lstlisting}
\end{minipage}\\

Instances of \textit{Message with Exception Trace (MET)} happen when there is an exception trace displayed in the app as a result of a connectivity problem.
 In \sfigref{sfig:nimmetexample} we present an example of this behavior in AntennaPod \textit{(v1.6.2.3)}, showing a message saying ``An error ocurred: java.net.UnknownHostException: Unable to resolve host "gpodder.net" [...]". As is can be seen in \listref{lst:nimmetAntennapod}, the message that is shown to the user is built using the caught exception.

\noindent
\begin{minipage}{\linewidth}
	\begin{lstlisting}[language={Java}, label={lst:nimmetAntennapod}, caption={Example of code for displaying Message with Exception Trace}, firstnumber=147]
txtvError.setText(getString(R.string.error_msg_prefix) + exception.getMessage());
	\end{lstlisting}
\end{minipage}\\

\textit{Inconsistent Message (IM)} groups the cases in which the app reports the execution of an action, which from the user's perspective is not related to the action triggered. In these cases, it is worth noting that despite performing the correct action, the message shown by the app suggests that another action was performed. For instance, Openshop.io \textit{(v1.2)} exhibits this issue (see \figref{sfig:nimimexampleOpenShop}). When a user accesses to the Terms section with Internet connection, it displays info related to the terms when buying clothes and accessories. Conversely, when the user attempts to perform the same action without Internet connection, the app states that the content (\ie the terms section) is empty (in the same place where the terms should be listed). This message is inconsistent because there is actually content that should be displayed and that, due to the lack of connection, cannot be retrieved.% and a new inaccurate content saying that ``Terms section is empty'' is displayed.

Another example is in AntennaPod \textit {(v1.6.2.3)} when a user tries to subscribe to a podcast without having connection. It is worth noting that AntennaPod downloads all media related to the podcast when the ``Subscribe" button is pressed. To this, AntennaPod creates an instance of \texttt{DownloadService}  (\listref{lst:nimimAntennaPod1}). Then, the background service starts downloading the data. During the download, the method \texttt{saveDownloadStatus} sets a boolean variable \texttt{createReport} in the case of lack of connectivity. Finally, this  variable is used to create a notification (\listref{lst:nimimAntennaPod2}). 

The notification displays the string \texttt{download\_report\_content} which is defined as ``\%1 \$d downloads succeeded, \%2\$d failed'' (\listref{lst:nimimAntennaPod3}). In summary, this scenario has an inconsistent message from the user's perspective because it is not informing  about the status of the ```Subscribe" request; the message reports the number of successful and failed downloads.

\noindent
\begin{minipage}{\linewidth}
	\begin{lstlisting}[language={Java}, label={lst:nimimAntennaPod1}, caption={Code snippet showing how a background service for downloading data is started in AntennaPod}, firstnumber=82]
Intent launchIntent = new Intent(context, DownloadService.class);
launchIntent.putExtra(DownloadService.EXTRA_REQUEST, request);
context.startService(launchIntent);
	\end{lstlisting}
\end{minipage}

\noindent
\begin{minipage}{\linewidth}
	\begin{lstlisting}[language={Java}, label={lst:nimimAntennaPod2}, caption={Code snippet for generating a notification in AntennaPod}, firstnumber=501]
if (createReport) {
	Log.d(TAG, "Creating report");
	// create notification object
	NotificationCompat.Builder builder = new NotificationCompat.Builder(this)
		.setTicker(getString(R.string.download_report_title))
		.setContentTitle(getString(R.string.download_report_content_title))
		.setContentText(
		String.format(
		getString(R.string.download_report_content),
		successfulDownloads, failedDownloads)
		)
	\end{lstlisting}
\end{minipage}\\
\begin{minipage}{\linewidth}
	\begin{lstlisting}[language={Java}, label={lst:nimimAntennaPod3}, caption={Inconsistent message definition at AntennaPod}, firstnumber=212]
<string name="download_report_content">%1$d downloads succeeded, %2$d failed</string>
	\end{lstlisting}
\end{minipage}\\
\subsection{Hybrid practices}
	
\emph{Hybrid practices} represent scenarios in which both good and bad practices are applied by developers. An example of these practices is when the user is trying to send an e-mail without Internet connection and two messages are displayed at the same time: One of the messages says ``An error occurred" and the other message states ``Your e-mail cannot be sent. It is going to be saved in the draft folder.". Note that both messages should appear separately to be considered as a hybrid practice. In this situation, there is a generic message about an error, but at the same time, there is a clear message about what is going to happen with the email because of the lack of Internet connection. This example of a hybrid practice is called \textit{Non-Informative Message \& Expressive Message (NIM-EM)}.

\begin{figure}[h!]
	\centering
	\subfigure[Browser Embedded Incorrectly but with an Expressive Message]
	{
		\label{sfig:beiemexample}
		\includegraphics[width=0.33\linewidth]{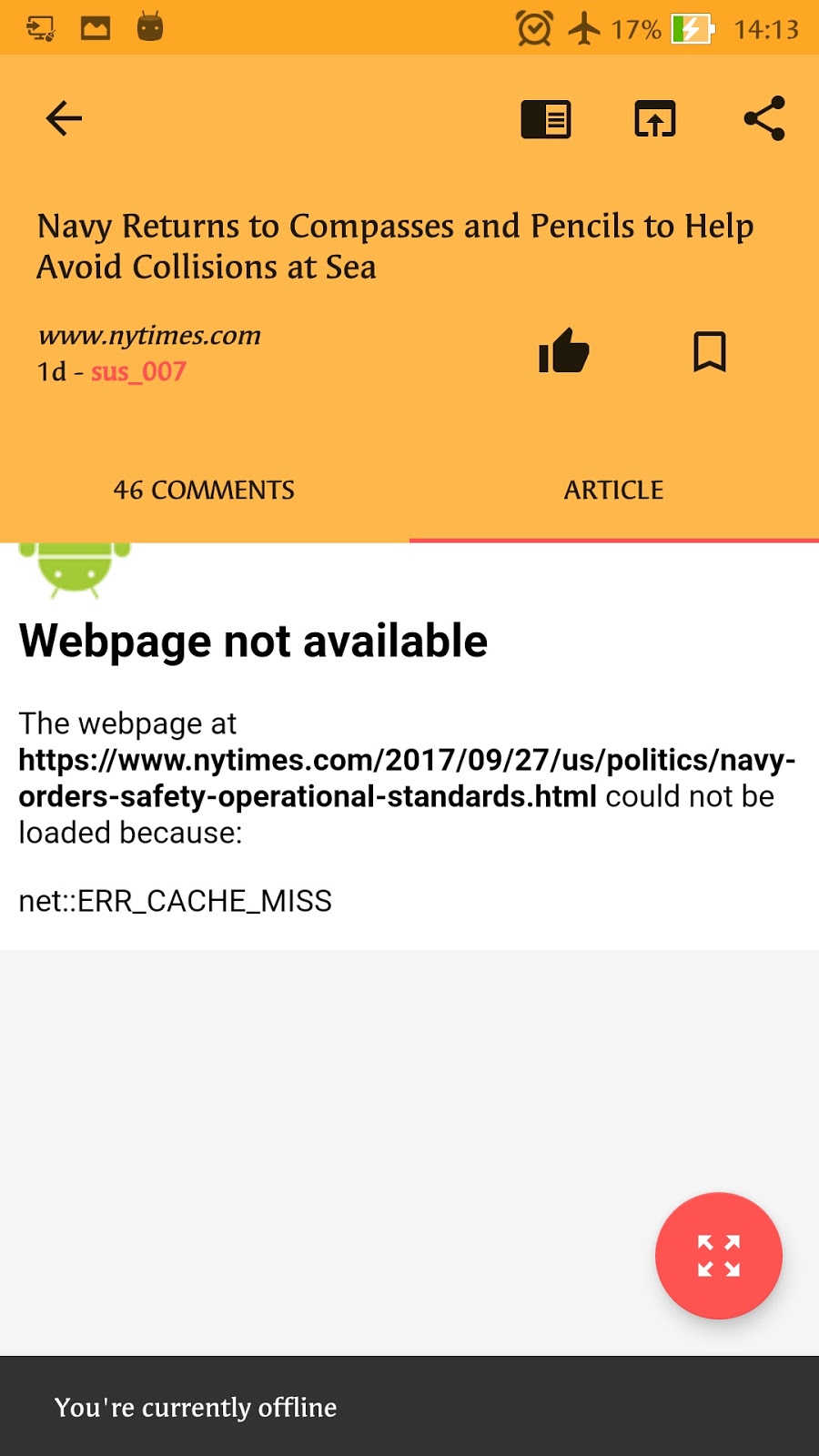}
	}
	\subfigure[Inexpressive Message plus Expressive Message]
	{
		\label{sfig:iememexample}
		\includegraphics[width=0.33\linewidth]{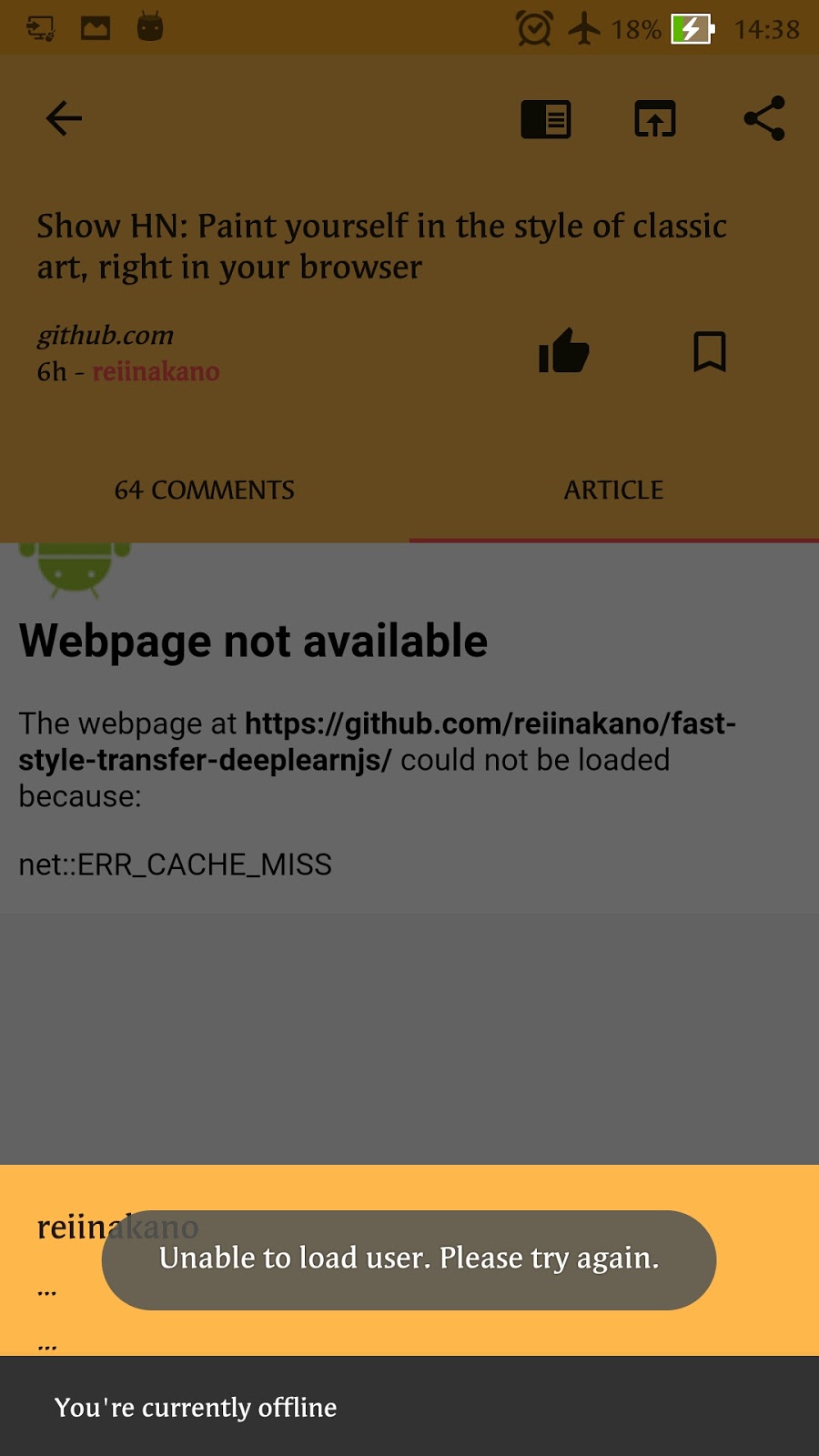}
	}
	\caption
	{
		Occurrences of Hybrid Patterns in Studied Applications 
	}
	\label{fig:ohp1}
\end{figure}  

In this section we present the hybrid practices identified in the analyzed apps.  In order to be clear which issue and good practice are happening, the hybrid practices are named using names referring to the observed issue and good practice, using the issue name first in order to emphasize the error. Each hybrid practice has its description and a label which represents it. It is worth noting that these are less common and we found only 4 instances in the ``Materialistic - Hacker News" mobile app \textit{(v3.1)}.	

\subsubsection{Browser Embedded Incorrectly \& Expressive Message (BEI-EM)}
In this case, the application shows an activity with an embedded browser displaying the default content for connectivity errors, and at the same time it displays an error message instead of the expected content. We consider this case as ``Browser Embedded Incorrectly'' because when there is a connectivity problem, the embedded browser should not be displayed or should not display the default error page (which is not expressive). This example was found in the  Materialistic - Hacker news app when a user selects a news from a list (this action triggers the display of the content associated with the news). However, for this purpose the app uses an activity with a header section where some metadata is shown (\eg News title, username of the contributor, etc.) and a body section where the url from the news is loaded using a WebView component (\ie embedded browser). Despite the error, after the load event is triggered by the WebView, the app makes a validation of the connection state in order to show a message. This message, even being an expressive message, is shown at the same time of the load message, as it can be seen in \sfigref{sfig:beiemexample}. Therefore, the user sees two messages that do not match in their meaning. 

The complete example at code level can be seen in our online appendix \cite{onlineAppendix}. However, the main code snippets are shown here. First, the starting point of this behavior as is can be seen in \listref{lst:beiemMaterialistic} happens when application call ``bindData'' method at line 177 and then validates the connection to show a message.

\noindent
\begin{minipage}{\linewidth}
	\begin{lstlisting}[language={Java}, label={lst:beiemMaterialistic}, caption={Browser Embedded Incorrectly but with an Expressive Message Code Snippet in Materialistic app}, firstnumber=176]
if (mItem != null) {
	bindData(mItem);
} else if (!TextUtils.isEmpty(mItemId)) {
	mItemManager.getItem(mItemId,
		getIntent().getIntExtra(EXTRA_CACHE_MODE, ItemManager.MODE_DEFAULT),
		new ItemResponseListener(this));
}
if (!AppUtils.hasConnection(this)) {
	Snackbar.make(mCoordinatorLayout, R.string.offline_notice, Snackbar.LENGTH_LONG).show();
}
	\end{lstlisting}
\end{minipage}\\

When ``bindData'' method is triggered the webview that shows the news is started and launched as it can be seen in \listref{lst:beiemMaterialistic2} at lines 412 and 413.

\noindent
\begin{minipage}{\linewidth}
	\begin{lstlisting}[language={Java}, label={lst:beiemMaterialistic2}, caption={Browser Embedded Incorrectly but with an Expressive Message Code Snippet in Materialistic app}, firstnumber=407]
if (story.isStoryType() && mExternalBrowser && !hasText) {
	TextView buttonArticle = (TextView) findViewById(R.id.button_article);
	buttonArticle.setVisibility(View.VISIBLE);
	buttonArticle.setOnClickListener(v ->
		AppUtils.openWebUrlExternal(ItemActivity.this, 
			story, story.getUrl(), mCustomTabsDelegate.getSession()));
}
	\end{lstlisting}
\end{minipage}\\

\subsubsection{Non-Informative Message \& Expressive Message (NIM-EM)}
In this case the app shows at the same time multiple messages when performing an action, some of them are inexpressive messages while the rest are expressive ones. For example, the Materialistic - Hacker news app redirects the user to an activity that shows that the operation was not successful, but at the same time, it shows an expressive message confirming that the user is using the app without an internet connection.

The reported example shows up when after selecting a news from a list, the user clicks on the contributors username to see more information; this action displays the content associated with the contributor. Due to connectionless state the information can not be retrieved and the app shows a generic message that does not provide enough information about the problem. However, the app makes a validation of the connection state in order to show a message. This message even being an expressive message is shown at the same time of the inexpressive message \sfigref{sfig:iememexample}.
%!TEX root = ../main.tex
%%%%%%%%%%%%%%%%%%%%%%%%%%%%%%%%%%%%%%

\subsection{ECn issues impact on Quality attributes}

\begin{figure}[h!]
	\centering
		\includegraphics[width=\linewidth]{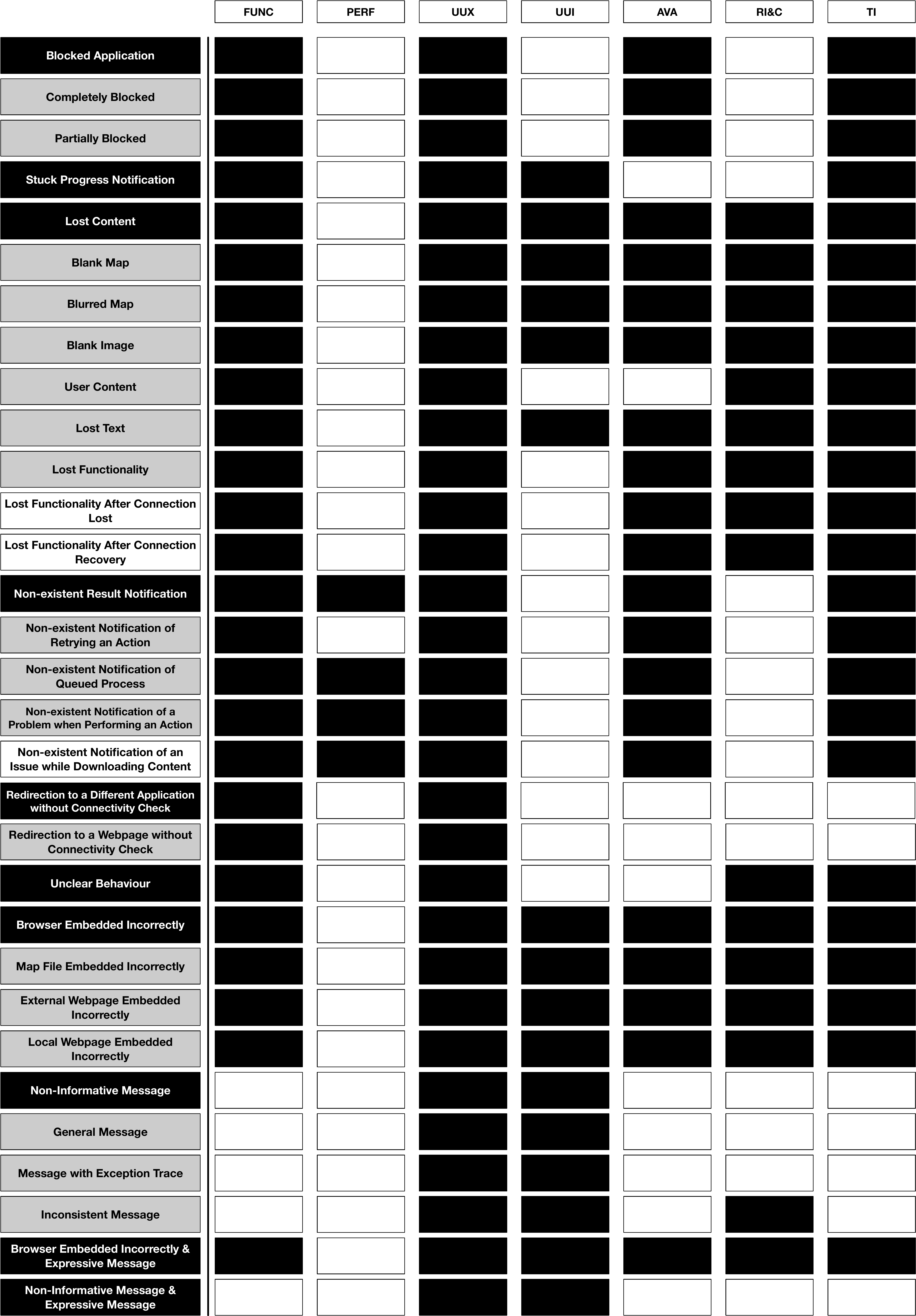}
		\vspace{-0.1cm}
		\caption{Impact of Eventual Connectivity bugs in Android apps to quality attributes. (FUNC) Functionality, (PERF) Performance, (UUX) User Experience, (UUI) User Interface Aesthetics, (AVA) Availability, (RI\&C) Resource integrity and Consistency, and (TI) Testability.}
		\label{fig:QAimpact}
\end{figure}

In addition to the taxonomy figure, we present in \figref{fig:QAimpact} how the ECn bug types identified in Android might potentially impact the set of quality attributes defined in \secref{sec:impFac}. In the figure, the columns represent quality attributes, and  the rows represent ECn issues. The colors associated to each issue type (\ie from white to black using different scales of gray) indicate their level in the taxonomy, with black being root categories and white the leaf categories. A black entry at the intersection between an issue type and a quality attribute indicates a possible \emph{negative impact} of the issue on the attribute, while a white entry represents \emph{no impact}. By \emph{negative  impact} we mean that the issue can affect the quality attribute as perceived by the user.

The starting point for defining the impact of the ECn issues was an open coding-inspired stage in which the first two authors, who are Ph.D. students with experience in mobile app development, assigned impact-level tags to each combination of issue type and quality attribute. Once the tags were independently assigned by the two authors, a merging stage was conducted to identify the cases in which there was disagreement: The two authors agreed on 189 out of the 217 entries, which represent the intersections of ECn issue types and quality attributes, resulting in a Krippendorff's alpha coefficient \cite{krippendorff2018content} indicating a substantial agreement level (0.73). All conflicts have then be solved through an open discussion.

Accordingly to our assessment, ECn bugs can thoroughly impact several quality characteristics within Android applications. From a quality attribute perspective, it is worth noticing that there are three main vertical clusters. Firstly, the most impacted quality attributes are User Experience (UUX) and User Interface Aesthetics (UUI). Therefore, when users face ECn issues in most cases they cannot successfully accomplish their tasks within the app, since (i) they cannot conduct such tasks neither in an effective nor efficient manner, and (ii) the interface does not enable a pleasing interaction in presence of ECn issues. A representative issue negatively impacting such quality attributes is \textit{Non-Informative Message}.

Secondly and consequently with the previous mentioned cluster, it is important to mention that ECn issues widely impact the Functionality (FUN) of the apps; this happens, for example, when an application is completely blocked, thus, the application does not provide internal functionalities meeting the needs of a user when experiencing lack of connectivity. Thirdly and in regards to Testability (TI), when apps exhibit ECn issues, some tests cannot be easily performed because complete test sequences can not be executed; therefore, code coverage and fault detection capabilities can be limited if  ECn issues do not allow test cases to explore and execute features of an app under test. 

The ECn issues having a negative impact on the widest range of quality attributes is \textit{Lost Content}, particularly the ones related to \textit{Lost Functionality}.

\section{Related work}
\label{section:relatedwork}

\emph{Network permission} is the most popular permission among Android apps \cite{Frank2012, Mostafa2017} which suggest that Android apps are highly  dependent on internet access. This has pushed the research community to study network-related aspects of mobile apps, and in particular: (i) network traffic profiling/characterization/usage \cite{Falaki2010, rao2011, Baghel2012, Ham2012,Wei2012b,Fukuda2015,Nayam2016,Mostafa2017,Conti2018,Rapoport2017}; (ii) security and privacy concerns, namely how private information is being manipulated and protected in apps \cite{Kuzuno2013, Song2015,Ren2016,Cheng2017,continella17,Huang2019,Gadient2020}; and (iii) network-related vulnerable scenarios which can be exploited by mischievous attackers \cite{Shabtai2011, Crussell2014,Wei2012a}. To the best of our knowledge, our study is the first one empirically analyzing \textit{\acl{ECn}} issues in Android apps. From this perspective, our work is mostly related to previous studies looking for other types of issues that can affect mobile apps. Thus, we start by discussing these works categorized by type of issue they investigate and, then, we briefly describe tools presented in the literature that, while not directly related to \textit{\acl{ECn}} issues, could be used to identify network-related issues.
 
%--------------------------------------------------------

% Head 2
\textbf{Security and privacy.}  Security-related issues affecting mobile apps are the most studied ``quality issues''. For this reason, we only focus on some representative works, since our purpose is not to present a complete survey of the literature in this field\footnote{The interested reader can refer to \cite{Sadeghi:TSE17,LI:IST17}.}. 

Munaiah \etal \cite{Munaiah:WAMA16} mined and reverse engineered $\sim$65k Android apps from the Google Play store in order to (i) classify them as malicious or benign, and (ii) collect a variety of quality and security-related information that have been organized in a dataset which could be used for further research in the field. 

Thomas \etal \cite{Thomas:SPSM2015} mined OS updates installed on 20k+ Android devices to measure the delivery time of security updates and to define a scoring model of insecure devices; their main finding suggests that, on average, 87.7\% of the devices are exposed to at least one of the 11 analyzed vulnerabilities. Moreover, Thomas \cite{Thomas:2015} investigated the CVE-2012-6636 \cite{CVE-2012-6636} vulnerability on the JavaScript-to-Java interface of the WebView API. The authors statically analyzed over 102k+ APKs in order to quantify the number of apps in which the security issue could be exploited, showing that after $\sim$5 years from the release of its security patch, the vulnerability was still exploitable in many apps. 

Linares-V\'asquez \etal \cite{Linares-Vasquez:MSR17} analyzed a set of 660 Android OS vulnerabilities (mined from CVE \cite{CVE}) to analyze their survivability, the subsystems and components of the Android OS involved in the vulnerabilities, and defined a taxonomy of security issues based on the Common Weakness Enumeration (CWE) hierarchy of vulnerabilities \cite{CWE}. This paper was later extended with a larger dataset by Mazuera-Rozo \etal \cite{mazuera2019vulns}.
 
Although our study also aims at identifying quality issues in Android apps as done in \cite{Munaiah:WAMA16, Thomas:2015}, we focus on eventual connectivity issues rather than security concerns. We share with the work by Linares-V\'asquez \etal \cite{Linares-Vasquez:MSR17} and Mazuera-Rozo \etal \cite{mazuera2019vulns} the manual analysis aimed at building a taxonomy of (different types of) issues. However, our taxonomy is not built by mining issues fixed by developers in open source projects, but rather by testing several  Android applications in order to identify connectivity issues in-the-wild.

\textbf{Performance in mobile apps.} Performance bugs are particularly relevant in mobile apps due to the limited resources usually available on devices running them (\eg limited energy available in their battery). For this reason, several studies have been conducted in this field. Lin \etal \cite{Lin:2014:RCA:2635868.2635903} performed a study on 104 popular open-source Android apps to investigate threading issues. This study provides evidence that even though half of the apps use AsyncTask, there is a huge number of places where long-running operations are not encapsulated in AsyncTask. Guo \etal \cite{Guo:2013:CDR:3107656.3107706} proposed an Android specific approach called Relda, which has a special focus on characterizing and detecting resource leaks related to operations with hardware-components such as sensors. 

Linares-V\'asquez \etal \cite{7332486} presented a taxonomy of practices and tools adopted by developers to detect and fix performance bottlenecks. The authors surveyed 485 open source Android app and library developers. Also, they analyzed performance bugs and fixes in the app's repositories. Their findings show that Android developers rely on manual testing and analysis of the reviews for detecting performance bottlenecks. 

More similar to the design adopted in our study is the work by Liu \etal \cite{Liu:2014:CDP:2568225.2568229}. The authors conducted the first empirical study on performance bugs in  mobile apps by analyzing 70 real-world performance bugs collected from eight Android apps. As main contribution they manually classified the 70 bugs into three categories: (i) GUI lagging, (ii) energy leak, and (iii) memory bloat. 

Stemming from the aforementioned seminal paper Mazuera-Rozo \etal \cite{mazuera20androidperformance} aimed at expanding the empirical knowledge about performance bugs in mobile apps. The authors presented a larger study on the types of performance bugs affecting not only Android but also iOS applications. For each platform the authors manually analyzed 250 commits aimed at fixing real performance bugs, they categorized the type of bug being fixed and later the authors created two taxonomies of performance bugs for Android and iOS apps. 

In terms of goal, our work is similar to the latter, since we also aim at classifying in a taxonomy a specific type of issues affecting mobile apps (\ie \textit{\acl{ECn}} issues). However, the focus (\textit{\acl{ECn}} \emph{vs} performance issues) as well as the methodology of the two studies are different. While Mazuera-Rozo \etal analyzed commits in which developer fixed performance bugs, we manually executed and inspected 50 Android open source apps looking for \textit{\acl{ECn}} issues.

\textbf{Usability.} The usability of software interfaces has been tackled in the seminal work by Nielsen and Molich \cite{Nielsen1990}. With the advent of mobile platforms with device-specific constraints (\eg small screens) several researchers investigated usability issues that can affect mobile apps (an  exhaustive survey of the literature in this context can be found in \cite{Weichbroth2020,Canedo2019}). We discuss a few representative examples.

Ghazizadeh \etal \cite{Ghazizadeh2017} compare the usability of apps providing user guides in form of animations \emph{vs} text. To this aim the authors conduct a study in which two different versions of an Android application was provided to 68 users (one using animations and one using text). The authors found that users with animated guide spend less time to learn functionalities as compared to participants having a text-based guide. 

Parente Da Costa \etal \cite{Canedo2019} extend the work by Nielsen and Molich \cite{Nielsen1990} to mobile apps. These heuristics are derived through a systematic literature review, conveying a model intended to be used in empirical validation of the usability of mobile apps. Similarly, researchers and practitioners could use our taxonomy when investigating and validating how connectivity impacts user experience of mobiles apps. 

Bessghaier and Soui \cite{Bessghaier:AICCSA17} also conducted a study to measure usability of four Android hybrid apps based on a predefined list of 13 structural usability defects. The authors aimed at creating a usability defects base of examples of hybrid applications. Similarly, Lelli \etal \cite{Lelli2015} empirically identified and classified several types of GUI errors that can affect GUI, thus resulting in a GUI
fault model designed to categorize GUI faults. Escobar-Vel\'asquez \etal \cite{Escobar2020} present an empirical study on how internationalization can impact the GUIs of Android apps. In particular, the authors evaluated the changes, bugs and bad practices related to GUIs when strings of a given default language are translated to 7 different languages. They used a set of 31 Android apps and their translated versions for such purpose. While some of these studies are similar to our work for what concerns the study design (\ie manual inspection of issues in apps), the focus of our study is on a different category of issues.

\textbf{Energy issues.} Several studies have conducted research on energy issues in mobile apps \cite{cruz2019catalog,Pathak:MobiSys12,Pathak:HotNets11,Pathak:Eurosys12,Pathak:Eurosys11,DBLP:conf/msr/VasquezBBOPP14,Hao:GREENS12,Li:ISSTA13,Hao:ICSE13,Li:ISSTA14,Li:ICSME14,Wan:ICST15,Sahin:JSEP16,li:icse16,Babakol2020}. A representative work in the area is the one by Carroll and Heiser \cite{DBLP:conf/usenix/CarrollH10}, who performed a detailed analysis of the energy consumption of a smartphone, based on measurements of a physical device, thus presenting (i) a list of the different components of the device contributing to power consumption, and (ii) a model of the energy consumption for different usage scenarios. 

Cruz \etal \cite{cruz2019catalog} inspect commits, issues and pull requests of $\sim$1k Android and 756 iOS apps, then present a catalog of 22 design patterns that contribute to improve the energy efficiency of mobile applications. Linares-V\'asquez \etal \cite{DBLP:conf/msr/VasquezBBOPP14} analyze the energy consumption of APIs used in Android apps. For such purpose the authors measure energy consumption of method calls when executing typical usage scenarios in 55 mobile apps from different domains, summarizing their findings in a set of guidelines for developers and researches on how to reduce energy consumption while using APIs. The goal of our study is similar to the one by Linares-V\'asquez \etal \cite{DBLP:conf/msr/VasquezBBOPP14} (\ie release a set of guidelines for practitioners and academics) but in a different context.

\textbf{Detection of quality issues.}  The closest existing tools for automated detection of crashes related to eventual connectivity are CrashScope \cite{Moran:ICSE17,Moran:ICST16}, Thor \cite{Adamsen:ISSTA15} and the one proposed by Azim \etal \cite{Azim14}. These approaches systematically explore an app, generate events like intermittent network connectivity, and look for crashes (\ie the application stops). As the reader appreciated in our taxonomy, crashes because of lack of connectivity are only a subset of the issues taxonomy, that also include issues that impact usability and user experience.

In addition, JazzDroid \cite{Xiong2018} is an automated fuzzer for Android platforms taking a gray-box approach which injects into applications various environmental interference, such as network delays, thus identifying issues related, but not limited to, connectivity. More in general, there exist approaches to test and diagnose common poor responsiveness behaviours, such as Monkey \cite{MONKEY}, AppSPIN \cite{Zhao2019}, TRIANGLE \cite{Panizo2019} and the one proposed by Yang \etal \cite{Yang2013} can test Android applications in eventual connectivity scenarios. 
\section{Threats to Validity}
\label{sec:threats}

\textbf{Construct validity.} In our study, they are mainly related to the measurements we performed, and in particular, the subjectivity during the manual tagging and construction of the taxonomy.  The catalog of connectivity issues was extracted by manually testing Android apps and the testing was driven by execution scenarios defined by the authors accordingly to the guidelines described in \secref{sec:design-manual}. It is possible that some scenarios that lead to connectivity issues were not executed. However (i) we tried to be exhaustive when defining scenarios for any of the visible features requiring connectivity; (ii) the apps were tested manually to avoid any issue with automated tests (\eg inconsistent behaviour), and (iii) all the scenarios and corresponding information (including videos of the issues) are available in our online appendix \cite{onlineAppendix}. In the case of issues tagging, we mitigated the subjectivity bias with weekly meetings (involving four authors) to revise the defined tags. Three of the authors have experience in developing Android apps. When a new tag was reported or when the taxonomy changed (\eg because two tags were merged) the scenarios execution was redone to check whether the already executed scenarios generated behaviors representative of the new tags. The tags, description of the tags, taxonomy, and mapping between scenarios and tags were always visible to the taggers to reduce the probability of duplicating tags.
Subjectivity is also a concern when it comes to the definition of ECn issues itself. Indeed, different app's users can perceive as more/less problematic specific types of ECn issues defined in our taxonomy. In our work, we considered as an ECn issue anything that could negatively affect the user experience in eventual connectivity scenarios. Assessing the relevance of the issues we identified with developers is part of our future research agenda. Finally, a threat could be generated due to the usage of real devices, since background tasks can impact the evaluation of the scenarios and using different devices could lead to have different execution conditions between taggers.

\textbf{Internal validity.} We are aware that external factors we did not control could affect the apps execution and results. To mitigate the effect of those factors we designed detailed scenarios that describe the device, app and their version, connectivity states, among other information (see \secref{sec:design-manual}). Therefore, the scenarios could be replicated with the same conditions. In addition, the scenarios execution were recorded and screenshots of the results were collected (see our online appendix \cite{onlineAppendix}).

\textbf{External validity.} We analyzed 50 open source native Android apps. We focused on open source apps because we wanted to analyze source code looking for implementation errors leading to the connectivity issues. Therefore, we recognize that we can not generalize our results to commercial/non-free apps that could follow different implementation and testing practices. Our target of 50 apps is another threat to external validity. However, we found such a number to be a good compromise between the generalizability of our study and the effort required for the data collection. Indeed, for each app we had to build the scenarios, manually execute them and manually analyze the source code looking for the root cause of the observed issue. Such a process required six months of work. While we do not claim generalizability of our findings over the set of apps we analyzed, we targeted heterogeneity of the considered apps that belong to 20 different domain categories, and have a medium-to-high-quality as perceived by users and reported in review ratings (min=3.4, median=4.3, max=4.8). Nevertheless, we acknowledge that our findings cannot be generalized outside of the set of 50 studied apps.

It is also important to highlight our focus on connectivity issues related to the Internet connection rather than to other network protocols (e.g., Bluetooth, NFC). Such an analysis, that would require the usage of multiple devices rather than the single one we used, is part of our future work.

Finally, it is worth mentioning that, due to our selection criteria we focused on apps having an average rating above 3.0. This means that the derived taxonomy might not be fully representative of apps having a substantially lower rating that could be affected by a more diverse set of ECn issues, possibly even more severe. Additional studies are needed to investigate this aspect.

\textbf{Conclusion validity.} This type of threats concerns the relationship between treatment and outcome. This is an observational study and  we did not consider statistical tests because we were neither interested on comparing samples nor measuring significance of differences. However, we used exploratory data analysis techniques to understand the apps sample. We report frequencies for each of the identified issues (see \figref{fig:taxonomy}), and each scenario is mapped to their corresponding tags.
\section{Learned Lessons and Future Work}
\label{sec:conclusions}

We analyzed 50 popular open source apps looking for bad practices. We executed 971 scenarios designed to test app functionalities that rely on Internet connection, with different connectivity scenarios. As result we found 320 issues that we have grouped into 12 categories. Our study is relevant for both researches and practitioners: The former could use the findings in our paper to design approaches aimed at automatically identifying connectivity issues. For the latter, we propose a list of recommendations discussed in the following.

\subsection{Checking the Connection Status}

Most of the issues we found are related to a missing verification of the connection state before performing an action. This was the root cause for many of the errors and exceptions that, when not being properly handled, cause application blockage, execution break, and the lack of informative messages. The connection status can be obtained using the ``Connectivity Manager" \cite{androidConnectivityManager, manageNetworkUsage}. The correct verification of connection status can prevent several of the error types in our taxonomy. For example, the \textit{Blocked Application} issues could be avoided by not triggering a request in case the connection is not ready. Note that request time-outs can also lead to those issues, however, for this case a post-API invocation check  is required to avoid  issues, such  as try-catch blocks for catching  the corresponding exceptions, or \texttt{onErrror} method definitions when the network operation is  invoked with a Future/Promise. If practitioners prefer to monitor the  connectivity  status while the  app is  running, it can be  done by using a \texttt{BroadcastReceiver} as described in \cite{monitorNetworkConnectivity}.
 
Another example is the Map Component that belongs to Google Maps. It uses an API in order to display a map as a Fragment. However, it does not provide a mechanism to handle the lack of connectivity when retrieving map information. This results in showing a \textit{black map} when using this component without connectivity. We also found several cases in which accessing a WebView component without checking the connection status resulted in errors. 

\subsection{Proper Handling of Libraries}

\subsubsection{Correct use of callbacks}

Most of the libraries that provide HTTP services use callbacks function. It is important to use wisely the error callback to improve the user experience. Additionally, if the backend service is not managed by the app developer, it is important to fully understand the default values used by the library as response when no result is found (\eg due to missing connectivity). Therefore, if there is an error or the response is the library's default value, the app could react properly without execution breaks. 

\subsubsection{Thread Handling}

When using libraries to connect to a backend service it is important to understand how responsibilities are delegated. Some of the libraries delegate the management of threads to developers. Also, heavy processes must be performed in a worker/secondary thread (rather than in the main one handling the user interface) to improve the user experience and avoid GUI lagging and ANRs \cite{7332486}\footnote{Note that Androids apps are prone to GUI lags and  ANRs because  of the single thread policy of the Android framework. }. In  the case of Kotlin, coroutines can be used with  an specific dispatcher that is optimized for network operations off the main thread (\ie \texttt{Dispatchers.IO)})\cite{kotlinIO,coroutinesADV}. In the case of Java-Android apps, worker threads for network operations can  be implemented by using  \texttt{Executor}\cite{runningBackground}, \texttt{Handler} \cite{runningBackgroundHandles}, the deprecated \texttt{AsyncTask}\cite{asyncTaskClass}, or the classic Java Threads.  Finally, off-the-main-thread HTTP requests are automatically handled by libraries like Volley\cite{volley} and Retrofit with asynchronous calls \cite{retrofit}.

\subsubsection{Map Libraries}

Using the GoogleMaps component is not the only way to display maps in Android apps. Other libraries allow developers to improve the map experience, applying custom themes and layers. However, it is important to be aware of possible issues these libraries could have, such as those documented in our study. Even more important, it is crucial for apps to properly inform the user about what is being shown. For example, if an app downloads a shell map that is used as default map when there is no connection, the user must be aware of this and knowing the specific features (\eg zooming) will not be available.

\subsection{Considering the Behavior of Android Components}

Knowing the behavior of the Android components is fundamental to properly handle cases in which connectivity errors arise. For example, the \texttt{SwipeRefreshLayout} component allows the user to use the ``Swipe-down" gesture to refresh the GUI. Such a behavior is managed by overwriting the \texttt{onRefresh()} method that is called each time the user ``refreshes" the GUI. While the application processes the refresh action, the \texttt{SwipeRefreshLayout} displays a progress notification that hides only once \texttt{setRefreshing(false)}  is invoked. Therefore, if the process in \texttt{onRefresh()} fails and the \texttt{setRefreshing(false)} method is not called, the progress notification will stay on screen until a further (successful) refresh is performed.

\subsection{Proper use of Informative Messages and Notifications}

One of the most common issue we found is related to messages not providing the correct information to the user. These issues can be avoided by adopting the following practices:

\begin{itemize}
	\item Exception Messages, unless properly phrased by the app developer, must not be included in the messages shown to the user. Most of the times users do not know the meaning of message like: \texttt{IndexOutOfBoundsException}, \texttt{NullPointerException}, \\texttt{ConnectException}, etc".
	\item The phrasing of the messages must take into account their ``consumer'' (\ie the app's user): It is important to avoid  technical words (\eg server, exception, connection error, \etc). 
	\item Avoid reusing messages in several parts of the application: They could lead to misunderstandings due to the fact those messages are, more often than not, quite generic and may not provide enough information for users to understand the problem.
\end{itemize}

\begin{figure}[h!]
 	\centering
 	\subfigure[Spotify]
 	{
 		\includegraphics[width=0.22\linewidth]{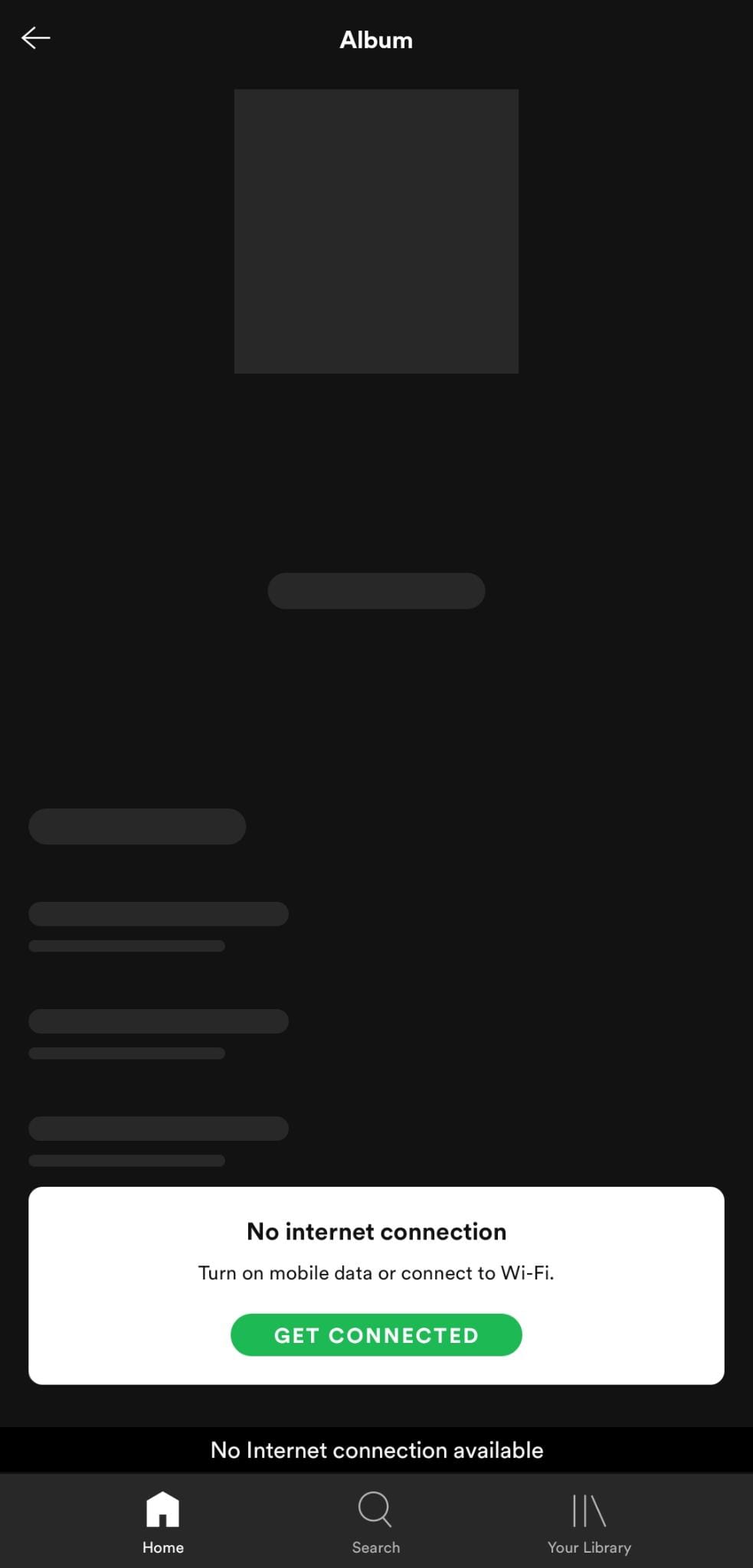}
 	}
 	\subfigure[Google Play Store]
 	{
 		\includegraphics[width=0.22\linewidth]{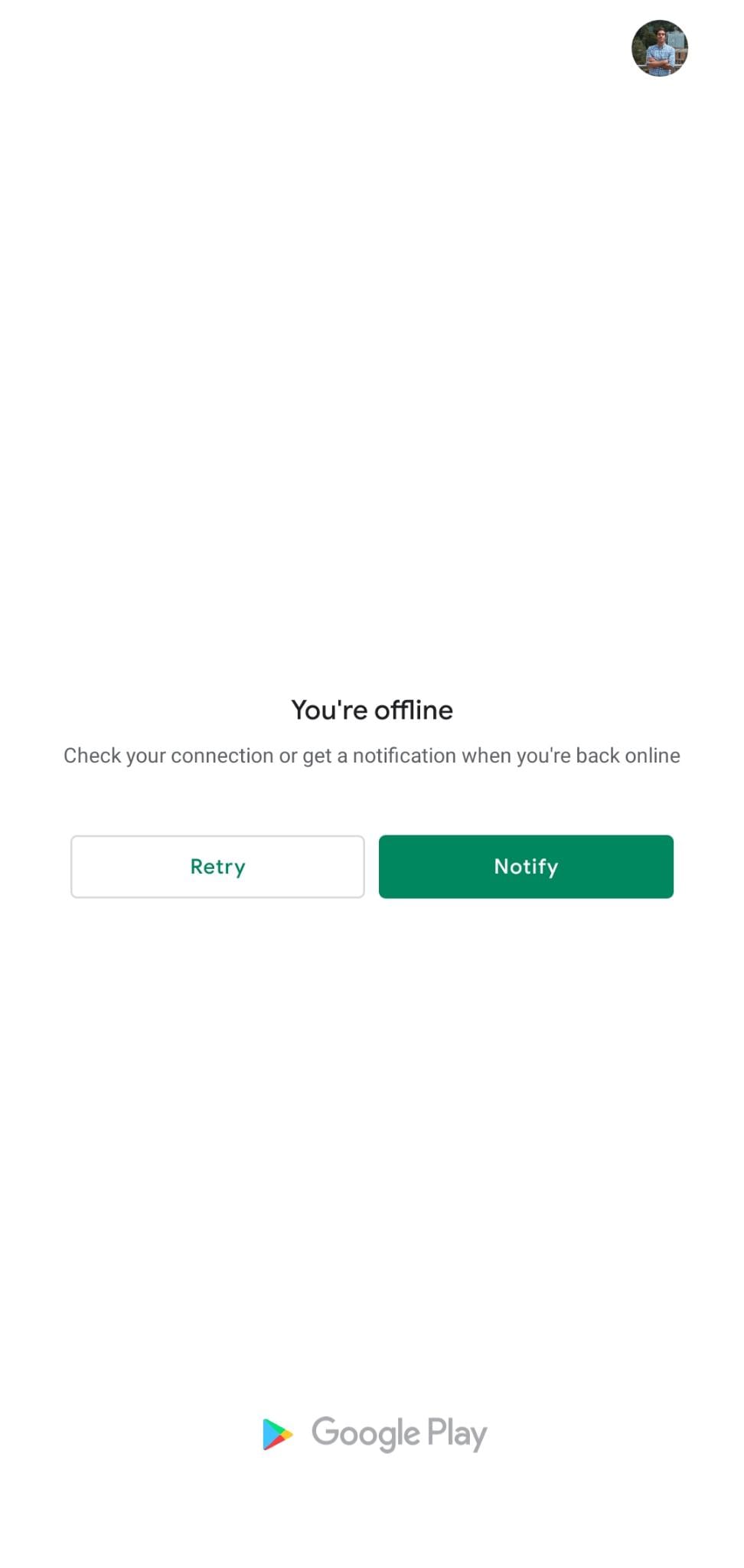}
 	}
	\subfigure[TikTok]
	{
		\includegraphics[width=0.22\linewidth]{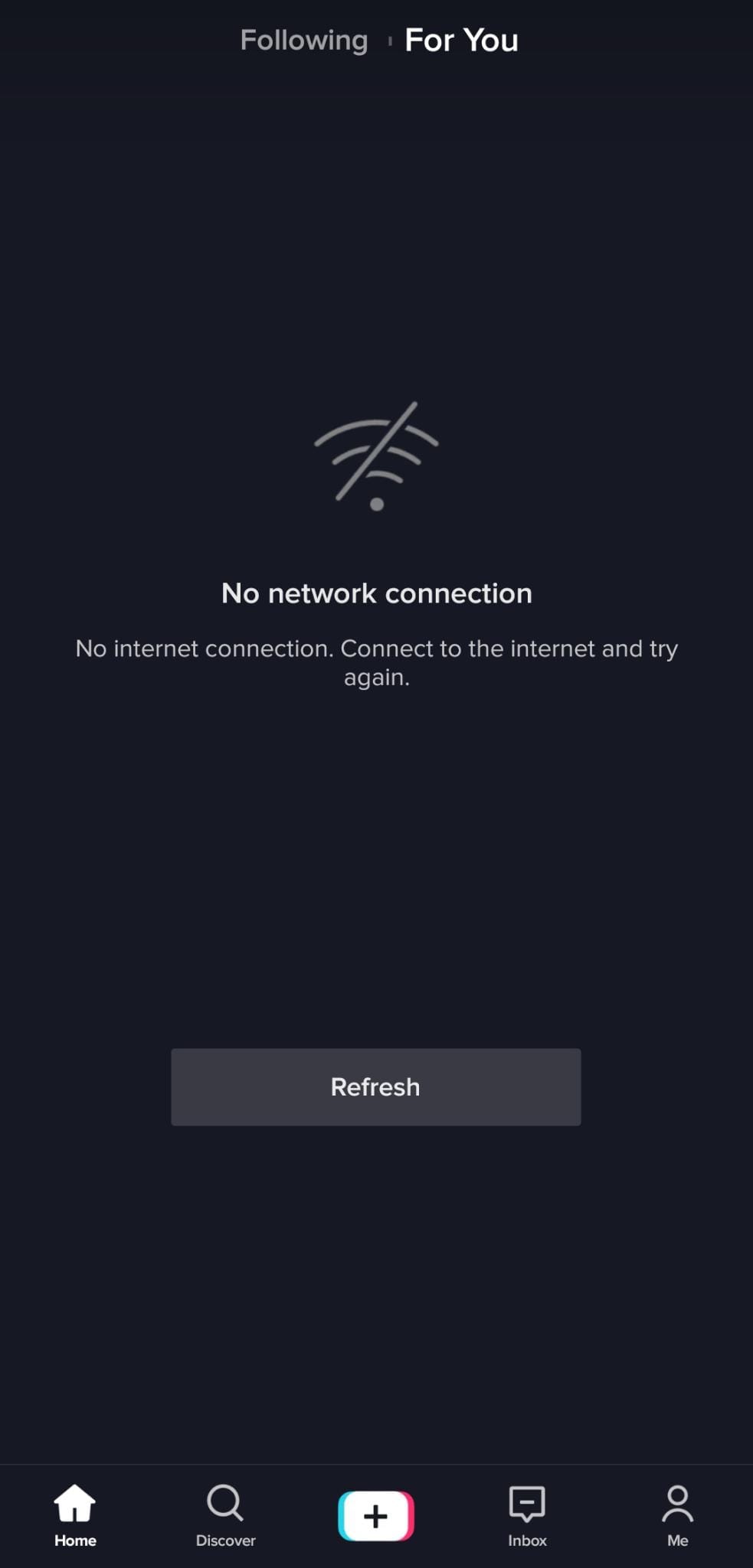}
	}
	\subfigure[Youtube]
	{
		\includegraphics[width=0.22\linewidth]{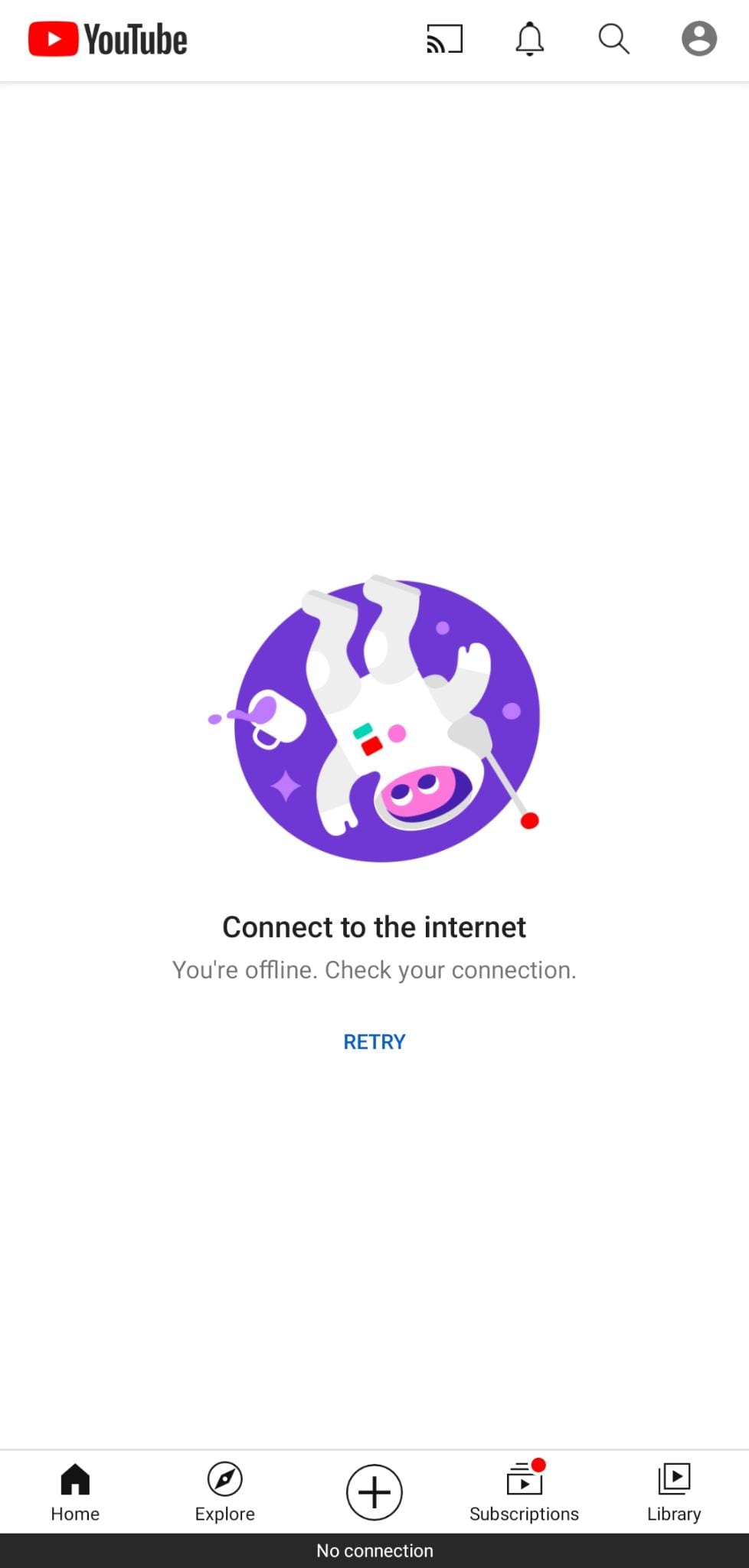}
	}
 	\caption
 	{
 		Examples of  expressive messages reporting connectivity issues 
 	}
 	\label{fig:goodExamples}
 \end{figure}

Similarly, a proper use of notifications is required. As recommended in the Android Developers Guide (ADvG), notifications should be used for foreground services and, in general, for notifying users of events relevant for them (\eg a request sent to an online service, the progress of such a request, \etc). A proper combination of notifications and meaningful messages can substantially improve the user experience. \fref{fig:goodExamples} illustrates examples of this  notifications in  Android apps that follow good practices for notifying users about connectivity issues.
 
\subsection{Queuing Actions Managing User Generated Content}

As shown in our study, connectivity issues can result in the lost of content or actions generated by users. To avoid this, all user-generated content and  actions that require network operations (\eg sending  a message in a communications apps) must be queued or locally stored when connection is  not available. In this way, in case of connectivity errors, the content and  actions will stay in a queue/cache  and can be triggered/submitted again when connectivity is back. It is also important to show to the user that its content has not been sent/processed yet, but it is stored for future actions. This makes the user aware of both the connection problem and the ``safe'' state of its data. To enable this, offline-first strategies \cite{offlineCookbook} must be implemented by relying on a combination of cached/local data and an actions queue.

Local caching can be implemented in Android apps by  using:
\begin{itemize} 
\item Application specific on-memory data structures
\item Android specific data structures such as the \texttt{LruCache} \cite{lruCache}
\item Retrofit and Volley  to enable HTTP requests caching 
\item the Picasso\cite{picasso} and Glide\cite{glide} libraries for image caching
\item  Firebase framework \cite{firebaseAccessOffline} or Realm \cite{realm} to enable caching of non-relational collection
\item SQLite for local storage of relational data
\end{itemize}

To enable a local actions queue, the \texttt{WorkManager} \cite{workmanager} and \texttt{RequestQueue} \cite{requestQueue} APIs can be used. In addition, as part of the JetPack architecture components, Google suggest the usage of the \texttt{Repository} design pattern \cite{cacheData} that allows retrieval of cached data when connection is not  available.

 Finally, a widely used strategy in web apps is  to display ``generic fallbacks''  when data is neither available from cache not network. A  fallback is a local resource (\eg image) that is displayed to the user  when the expected data is not available, with the purpose of improving the user experience or  avoiding layout issues.

\subsection{Delegating Actions to Other Apps}

When an app must perform an action already implemented in pre-installed apps (\eg sending an email), they can delegate the execution of such a feature. However, this means that the main application loses control of behavior and an error in the delegated app might lead the user to think that the main app has failed too. Knowing that the delegated actions require connectivity should push the app's developers to properly handle failing cases in order to communicate to their uses the issue experienced with the delegated app.

\subsection{Summing Up}
While we tried to summarize in  this  paper the most important lessons learned from our study, our online appendix \cite{onlineAppendix} provides developers with an extensive catalogue of ``bad-practices'' that must be avoided when handling eventual connectivity scenarios. Those bad-practices and the aforementioned recommendations could be used by  researchers  to design detectors and tools for automated refactoring/patching of eventual connectivity issues.

For instance, static analyses could be used to  detect code statements invoking network operations and then (i) identify the  existence of  connectivity status checks and  statements for catching connectivity errors such as time-outs or null responses, and  (ii) verify the proper execution of the network  operations  off-the-main-thread. Such analyses will help practitioners to mitigate several of  the reported issues such as \textit{BA}, \textit{LC}, and \textit{RDAC}. An interesting  avenue for  research here is  the automated detection of \textit{NERN} and  \textit{NIM}; this requires more specialized static analyses and automated test cases generation that are able to execute the features related to the network operations and check that  there are visual hints when connectivity  is disabled in  the device. Finally, recommender systems and tools for automated refactoring could be designed to analyze source code and suggest the usage of good implementation  practices (\eg generic fallback, repository pattern, request queue, images caching, etc), and  also automatically  apply the corresponding refactorings.

\begin{acknowledgements}
Escobar-Vel\'asquez and Linares-V\'asquez were partially funded by a Google Latin America Research Award 2018-2021. Escobar-Velásquez was supported by a ESKAS scholarship, No. 2020.0820. Mazuera-Rozo and Bavota gratefully acknowledge the financial support of the Swiss National Science Foundation for the CCQR project (SNF Project No. 175513).
\end{acknowledgements}

\bibliographystyle{IEEEtran}
\bibliography{local,bib/testing,bib/tools}

\end{document}